\documentclass[twocolumn,tighten]{aastex63}
\usepackage{natbib}

\received{2020-04-28}
\revised{2020-07-10}
\submitjournal{\apjs}
\accepted{2020-07-13}
\shorttitle{The SDSS Quasar Catalog: DR16Q}
\shortauthors{Lyke et al.\ 2020}

\newcommand{\kms}{$\textrm{km\,s}^{-1}$}
\newcommand{\hubu}{$\textrm{km\,s}^{-1}\,\textrm{Mpc}^{-1}$}
\newcommand{\micm}{$\mu\textrm{m}$}
\newcommand{\wmh}{$\textrm{W\,m}^{-2}\,\textrm{Hz}^{-1}$}
\newcommand{\kev}{\textrm{keV}}
\newcommand{\cm}{\textrm{cm}}
\newcommand{\mjy}{\textrm{mJy}}
\newcommand{\ecs}{$\textrm{erg\,s}^{-1}\,\textrm{cm}^{-2}$}
\newcommand{\esec}{$\textrm{erg\,s}^{-1}$}
\newcommand{\masy}{$\textrm{mas\,yr}^{-1}$}
\newcommand{\mm}{\textrm{mm}}
\newcommand{\degdeg}{$\textrm{deg}^{-2}$}
\newcommand{\degsqr}{$\textrm{deg}^{2}$}

\newcommand{\ciii}{\ion{C}{3}]} %C III at 1908 Ang
\newcommand{\civ}{\ion{C}{4}} %C IV at 1549 Ang
\newcommand{\ha}{H$\alpha$} %H alpha
\newcommand{\hb}{H$\beta$} %H beta
\newcommand{\hg}{H$\gamma$} %H gamma
\newcommand{\oii}{[\ion{O}{2}]} %O II at 3728
\newcommand{\oiii}{[\ion{O}{3}]} %O III at 5008
\newcommand{\lya}{Ly$\alpha$} %Lyman alpha
\newcommand{\mgii}{\ion{Mg}{2}} %Mg II at 2800
\newcommand{\siiv}{\ion{Si}{4}} %Si IV at 1400 Ang

\newcommand{\zpca}{$Z_{\textrm{\footnotesize{PCA}}}$}
\newcommand{\zpipe}{$Z_{\textrm{\footnotesize{pipe}}}$}
\newcommand{\zsys}{$Z_{\textrm{\footnotesize{HOST}}}$}

\newcommand{\zt}{$Z_{\textrm{\footnotesize{T}}}$}
\newcommand{\rvb}{\texttt{redvsblue}}
\newcommand{\qnet}{\texttt{QuasarNET}}

\begin{document}

\title{The Sloan Digital Sky Survey Quasar Catalog: Sixteenth Data Release}

\author[0000-0002-5078-2472]{Brad W.\ Lyke}
\affiliation{University of Wyoming, 1000 E.\ University Ave., Laramie, WY 82071, USA}
\correspondingauthor{Brad W.\ Lyke}
\email{blyke@uwyo.edu}

\author{Alexandra N.\ Higley}
\affiliation{University of Wyoming, 1000 E.\ University Ave., Laramie, WY 82071, USA}

\author{J.~N.\ McLane}
\affiliation{University of Wyoming, 1000 E.\ University Ave., Laramie, WY 82071, USA}

\author{Danielle P.\ Schurhammer}
\affiliation{University of Wyoming, 1000 E.\ University Ave., Laramie, WY 82071, USA}

\author{Adam D.\ Myers}
\affiliation{University of Wyoming, 1000 E.\ University Ave., Laramie, WY 82071, USA}

\author{Ashley J. Ross}
\affiliation{The Center of Cosmology and Astro-Particle Physics, Ohio State University, Columbus, Ohio, USA}

\author{Kyle Dawson}
\affiliation{Department of Physics and Astronomy, University of Utah, 115 S.\ 1400 E., Salt Lake City, UT 84112, USA}

\author{Sol{\`e}ne Chabanier}
\affiliation{IRFU, CEA, Centre d'Etudes Saclay, 91191 Gif-Sur-Yvette Cedex, France}

\author{Paul Martini}
\affiliation{The Center of Cosmology and Astro-Particle Physics, Ohio State University, Columbus, Ohio, USA}
\affiliation{Department of Astronomy, The Ohio State University, 140 W.\ 18th Ave., Columbus, OH 43210, USA}

\author{Nicol{\'a}s G.\ Busca}
\affiliation{Sorbonne Universit{\'e}, CNRS/IN2P3, Laboratoire de Physique Nucl{\'e}aire et de Hautes Energies,\\LPNHE, 4 Place Jussieu, 75252, Paris, France}

\author{H{\'e}lion du Mas des Bourboux}
\affiliation{Department of Physics and Astronomy, University of Utah, 115 S.\ 1400 E., Salt Lake City, UT 84112, USA}

\author{Mara Salvato}
\affiliation{Max-Planck-Institut f{\"u}r Extraterrestrische Physik, Giessenbachstra{\ss}e 1, 85748 Garching, Germany}

\author{Alina Streblyanska}
\affiliation{Instituto de Astrofisica de Canarias (IAC), E-38200 La Laguna, Tenerife, Spain}
\affiliation{Universidad de La Laguna (ULL), Dept. Astrofisica, E-38206 La Laguna, Tenerife, Spain}

\author{Pauline Zarrouk}
\affiliation{Institute for Computational Cosmology, Department of Physics, Durham University, South Road, Durham, DH1 3LE, U.K.}
\affiliation{IRFU, CEA, Centre d’Etudes Saclay, 91191 Gif-Sur-Yvette Cedex, France}

\author{Etienne Burtin}
\affiliation{IRFU, CEA, Centre d’Etudes Saclay, 91191 Gif-Sur-Yvette Cedex, France}

\author{Scott F.\ Anderson}
\affiliation{Department of Astronomy, University of Washington, Box 351580, Seattle, WA 98195, USA}

\author{Julian Bautista}
\affiliation{Institute of Cosmology \& Gravitation, University of Portsmouth, Dennis Sciama Building, Portsmouth, PO1 3FX, UK}

\author{Dmitry Bizyaev}
\affiliation{Apache Point Observatory and New Mexico State University, P.O.\ Box 59, Sunspot, NM, 88349-0059, USA}
\affiliation{Sternberg Astronomical Institute, Moscow State University, Moscow, Russia}

\author{W.\ N.\ Brandt}
\affiliation{Department of Astronomy \& Astrophysics, 525 Davey Lab, The Pennsylvania State University, University Park, PA 16802, USA}
\affiliation{Institute for Gravitation and the Cosmos, The Pennsylvania State University, University Park, PA 16802, USA}
\affiliation{Department of Physics, 104 Davey Lab, The Pennsylvania State University, University Park, PA 16802, USA}

\author{Jonathan Brinkmann}
\affiliation{Apache Point Observatory and New Mexico State University, P.O.\ Box 59, Sunspot, NM, 88349-0059, USA}

\author{Joel R.\ Brownstein}
\affiliation{Department of Physics and Astronomy, University of Utah, 115 S.\ 1400 E., Salt Lake City, UT 84112, USA}

\author{Johan Comparat}
\affiliation{Max-Planck-Institut f{\"u}r Extraterrestrische Physik, Giessenbachstra{\ss}e 1, 85748 Garching, Germany}

\author{Paul Green}
\affiliation{Harvard-Smithsonian Center for Astrophysics, 60 Garden St.\, MS \#20, Cambridge, MA 02138, USA}

\author{Axel de la Macorra}
\affiliation{Instituto de Física, Universidad Nacional Aut{\'o}noma de M{\'e}xico, Apdo.\ Postal 20-364, M{\'e}xico}

\author{Andrea Mu{\~n}oz Guti{\'e}rrez}
\affiliation{Instituto de Física, Universidad Nacional Aut{\'o}noma de M{\'e}xico, Apdo.\ Postal 20-364, M{\'e}xico}

\author{Jiamin Hou}
\affiliation{Max-Planck-Institut f{\"u}r Extraterrestrische Physik, Giessenbachstra{\ss}e 1, 85748 Garching, Germany}

\author{Jeffrey A.\ Newman}
\affiliation{PITT PACC, Department of Physics and Astronomy, University of Pittsburgh, Pittsburgh, PA 15260, USA}

\author{Nathalie Palanque-Delabrouille}
\affiliation{IRFU, CEA, Centre d’Etudes Saclay, 91191 Gif-Sur-Yvette Cedex, France}

\author{Isabelle P{\^a}ris}
\noaffiliation{}

\author{Will J.\ Percival}
\affiliation{Waterloo Centre for Astrophysics, University of Waterloo, 200 University Ave W, Waterloo, ON N2L 3G1, Canada}
\affiliation{Department of Physics and Astronomy, University of Waterloo, 200 University Ave W, Waterloo, ON N2L 3G1, Canada}
\affiliation{Perimeter Institute for Theoretical Physics, 31 Caroline St.\ North, Waterloo, ON N2L 2Y5, Canada}

\author{Patrick Petitjean}
\affiliation{Institut d'Astrophysique de Paris, Sorbonne Universit{\'es} and CNRS, 98bis Boulevard Arago, 75014 Paris, France}

\author{James Rich}
\affiliation{IRFU, CEA, Centre d’Etudes Saclay, 91191 Gif-Sur-Yvette Cedex, France}

\author{Graziano Rossi}
\affiliation{Department of Astronomy and Space Science, Sejong University, 209, Neungdong-ro, Gwangjin-gu, Seoul, South Korea}

\author{Donald P.\ Schneider}
\affiliation{Department of Astronomy and Astrophysics, The Pennsylvania State University, University Park, PA 16802}
\affiliation{Institute for Gravitation and the Cosmos, The Pennsylvania State University, University Park, PA 16802}

\author{Alexander Smith}
\affiliation{IRFU, CEA, Centre d’Etudes Saclay, 91191 Gif-Sur-Yvette Cedex, France}

\author{M.\ Vivek}
\affiliation{Indian Institute of Astrophysics, Koramangala, Bangalore 560034, India}
\affiliation{525, Davey Lab, Pennsylvania State University, State College, PA-16802, USA}

\author{Benjamin Alan Weaver}
\affiliation{NSF's National Optical-Infrared Astronomy Research Laboratory, 950 North Cherry Avenue, Tucson, AZ 85719, USA}

\begin{abstract}
    We present the final Sloan Digital Sky Survey IV (SDSS-IV) quasar catalog from Data Release 16 of the extended Baryon Oscillation Spectroscopic Survey (eBOSS). This catalog comprises the largest selection of spectroscopically confirmed quasars to date. The full catalog includes two sub-catalogs\footnote{The current versions are DR16Q\_v4 and DR16Q\_Superset\_v3 at: \url{https://data.sdss.org/sas/dr16/eboss/qso/DR16Q/}}: a ``superset'' of all SDSS-IV/eBOSS objects targeted as quasars containing 1{,}440{,}615 observations and a quasar-only catalog containing 750{,}414 quasars, including 225{,}082 new quasars appearing in an SDSS data release for the first time, as well as known quasars from SDSS-I/II/III. We present automated identification and redshift information for these quasars alongside data from visual inspections for 320{,}161 spectra. The quasar-only catalog is estimated to be 99.8\% complete with 0.3\%--1.3\% contamination. Automated and visual inspection redshifts are supplemented by redshifts derived via principal component analysis and emission lines. We include emission line redshifts for \ha, \hb, \mgii, \ciii, \civ, and \lya. Identification and key characteristics generated by automated algorithms are presented for 99{,}856 Broad Absorption Line quasars and 35{,}686 Damped Lyman Alpha quasars. In addition to SDSS photometric data, we also present multi-wavelength data for quasars from GALEX, UKIDSS, WISE, FIRST, ROSAT/2RXS, XMM-Newton, and Gaia. Calibrated digital optical spectra for these quasars can be obtained from the SDSS Science Archive Server.

\end{abstract}

\section{Introduction}\label{sec:intro}
The Sloan Digital Sky Survey (SDSS) has a long history of creating and releasing catalogs of quasars for use in cosmology and studies of quasar physics. SDSS spectroscopically observed 105{,}783 quasars released between the first \citep{schneider2002} and final \citep{dr7q_paper} quasar catalogs of the SDSS legacy programs (SDSS-I/II). When the third iteration of SDSS observations \citep[SDSS-III;][]{sdss3_mission} began, the combination of a new spectroscopic instrument, a new focus on quasar programs, extra fibers for spectroscopy, and the discontinuation of imaging observations, significantly increased the number of quasars that were observed. Over the course of the SDSS Baryon Oscillation Spectroscopic Survey \citep[BOSS;][]{boss_mission} program, three quasar catalogs were released: DR9Q, DR10Q, and DR12Q \citep{dr9q_paper,dr10q_paper,dr12q_paper}, with the final BOSS quasar catalog, DR12Q, containing 297{,}301 quasars. DR12Q also contained data from external catalogs in other wavelength ranges, multiple redshift estimates, and broad emission line parameters. DR14Q \citep{dr14q_paper} was the first quasar catalog released as part of SDSS-IV quasar program, containing 526{,}356 quasars. In this paper, we release the final quasar catalog of the SDSS-IV quasar program, the extended Baryon Oscillation Spectroscopic Survey \citep[eBOSS;][]{eboss_mission}. This catalog, which we will refer to as DR16Q, contains 920{,}110 observations of 750{,}414 quasars. Comparisons of DR16Q to previous quasar catalogs are limited to DR12Q and DR7Q, as these represent the final quasar catalogs accompanying each iteration of SDSS.

One of the primary objectives of eBOSS was to constrain the angular scale of baryon acoustic oscillations (BAO) in tracers of the distribution of matter \citep[for detection see][]{eisenstein2005,cole2005}. To achieve the planned precision of 2.8\% on the angular diameter distance, $D_{\rm A}(z)$, and 4.2\% on the Hubble Parameter, $H(z)$, at $z \sim 1.5$, eBOSS was designed to obtain spectra for a set of $\sim$500{,}000 quasars in the redshift range $0.8 < z < 2.2$ \citep{eboss_mission}. This represented more than a fivefold increase in the number of quasars for that redshift range, as compared to the final BOSS quasar catalog, DR12Q. With the final eBOSS observations complete, DR16Q comprises 480{,}459 quasars within the redshift range $0.8 < z < 2.2$. To constrain BAO measurements at higher redshift, eBOSS also planned to observe 120{,}000 quasars at $z > 2.1$, increasing the precision on $D_{\rm A}(z)$ and $H(z)$ from \lya\ forest measurements by a factor of 1.44. DR16Q includes 239{,}081 \lya\ quasars, a 25\% increase over DR12Q.

While the primary objectives of eBOSS and other SDSS-IV programs informed the target selection for quasars in DR16Q (see \S\ref{ssec:targeting}), eBOSS quasar catalogs have been used in a number of other recent research studies beyond the eBOSS core programs. These include, but are certainly not limited to: The study of changing-look quasars, which can be used to investigate accretion mechanisms and other quasar physics \citep{sheng2020}; X-ray studies of the clustering of quasars, which can also give insight into the growth and evolution of supermassive black holes \citep{powell2020}; studies of the correlation between X-ray and emission line luminosities \citep{timlin2020}; studies of quasar outflows in the far infrared and radio bands via follow-up of optically confirmed quasars \citep{hall2019}; studies of the correlation of outflow velocities with bolometric luminosity in Broad Absorption Line (BAL) quasars \citep{bruni2019}; and studies of the variability in BAL troughs over time \citep{grier2016,mcgraw2017}. To help facilitate these sorts of quasar studies, DR16Q includes multiple redshift estimates, BAL and damped \lya\ (DLA) quasar identifications, and compiled multi-wavelength data.

Previous iterations of SDSS quasar catalogs have included redshifts from automated classification pipelines, visual inspections, and principal component analysis (PCA) based on prominent quasar emission lines. In this tradition, DR16Q includes a range of redshift estimates that are characterized by different accuracies, precisions, and levels of homogeneity (see \S\ref{sec:redshifts} for more details). Most importantly, we have visually inspected 329{,}130 quasars in the catalog, with 326{,}535 of these having confident visual classifications and redshifts. Our set of visual inspections also includes the results from a random subsample of 10{,}000 eBOSS quasar targets. We visually inspected this random subsample and included inspections of different observations of the same quasar to characterize the accuracy and precision of  automated redshift algorithms that are also presented in this catalog (again see \S\ref{sec:redshifts} for more information).

This study is part of a coordinated release of the final eBOSS measurements of BAO and redshift space distortions (RSD) in the clustering of luminous red galaxies (LRG) \citep[$0.6<z<1.0$;][]{LRG_corr,gil-marin20a}, emission line galaxies (ELG) \citep[ $0.6<z<1.1$;][]{raichoor20a,tamone20a,demattia20a}, and quasars \citep[$0.8<z<2.2$;][]{hou20a,neveux20a}. An essential component of these studies is the construction of data catalogs \citep[this catalog;][]{ross20a}, mock catalogs \citep{lin20a,zhao20a}, and n-body simulations for assessing systematic errors \citep{rossi20a,smith20}. At the highest redshifts, $z > 2.1$, the coordinated release of final eBOSS measurements includes measurements of BAO in the \lya\ forest \citep{bourboux20a}. The cosmological interpretation of these results in combination with the final BOSS results and other probes is found in \citet{eBOSS_Cosmology}.

This paper is organized as follows: In \S\ref{sec:surveyoutline} we summarize the data used to target DR16Q quasars. In \S\ref{sec:construct} we outline how we constructed the two different catalog files released as part of DR16Q. In \S\ref{sec:redshifts} we describe the different redshift estimates in DR16Q, their uncertainties, and potential issues for each estimator. In \S\ref{sec:dlabal} we describe the automated algorithms used for identifying and analyzing Damped Lyman Alpha (DLA) systems and Broad Absorption Line (BAL) quasars. In \S\ref{sec:quasarsummary}, we outline the physical properties of the DR16Q quasar sample. In \S\ref{sec:multiwave} we briefly discuss the multi-wavelength data included in DR16Q. In \S\ref{sec:description} we describe the quantities included in the DR16Q quasar-only catalog, before concluding in \S\ref{sec:conclusion}. We also provide an Appendix that discusses the precision and accuracy of different redshift estimates for DR16Q quasars, presents some spectra of DR16Q quasars, notes mistakes that we have corrected from previous SDSS quasar catalogs, and details the data model of the DR16Q quasar-only catalog.

\section{Survey outline}\label{sec:surveyoutline}
In this section we summarize the imaging surveys, target selection procedures, and spectroscopic observations that produced the SDSS-IV/eBOSS quasar sample.

\subsection{Imaging Data for Targeting}\label{ssec:imaging}
Three sets of imaging data were used to generate quasar targets for SDSS-IV/eBOSS \citep{eBOSS_target_select}. The primary imaging was an updated calibration of SDSS-I/II/III. Additional imaging was incorporated from the Wide-field Infrared Survey Explorer \citep[WISE;][]{wise_mission_paper} using the custom ``unWISE'' coadds \citep{lang2014}. Finally, information on source variability from the Palomar Transient Factory \citep[PTF;][]{law2009,rau2009} was used to supplement quasar targeting \citep{palanque2016}.

SDSS imaging data, in the $u$, $g$, $r$, $i$, and $z$ photometric bands \citep{fukugita1996}, was taken at the 2.5\,m Sloan telescope \citep{gunn2006} using the 30 2k$\times$2k CCDs outlined in \citet{gunn1998}. By Data Release 8 \citep[DR8;][]{dr8_paper} over 14{,}000\,\degsqr\ of sky was covered by SDSS imaging. SDSS-IV/eBOSS used the same DR8 imaging as SDSS-III/BOSS, but leveraged new photometric calibrations using the ``uber-calibration'' method of \citet{uber_calib_method}, updated by \citet{schlafly_calib} to be pinned to PanSTARRS imaging \citep{panstarrs_paper}. A full description of this process can be found in \citet{finkbeiner2016}.

The WISE mission \citep{wise_mission_paper} collected data in four infrared bands: W1 (3.4\,\micm), W2 (4.6\,\micm), W3 (12\,\micm), and W4 (22\,\micm). \citet{unwise_coadds} used WISE data to create a custom set of coadded ``unWISE'' images, which were force-photometered at the locations of known SDSS sources by \citet{lang2014}. Due to significant differences in depth between W1/W2 and W3/W4 data, only W1 and W2 were used for eBOSS targeting.

PTF imaging was obtained in the Mould-\textit{R} filter and supplemented with SDSS $g$, as discussed in \citet{ofek2012}. For eBOSS targeting, a custom pipeline was used to coadd individual PTF frames on a timescale of 1 to 4 epochs per year. These coadded images were also used to construct a full stack, which was 50\% complete to known quasars at a magnitude limit of $\textit{g}\sim22.5$. A catalog of sources was extracted from this full stack and light curves were generated to supplement eBOSS targeting, as detailed in \citet{eBOSS_target_select}.

\subsection{Target Selection}\label{ssec:targeting}
One of the main goals of SDSS-IV/eBOSS was to study dark energy using the BAO method \citep[e.g.][]{ata2017}. Specifically, the eBOSS quasar sample was designed to achieve a precision on the angular diameter distance, $D_A(z)$, of 2.8\% for the redshift range $0.9 < z < 2.2$, and a 4.2\% precision in the value of $H(z)$ within that same redshift range for the quasar-quasar auto-correlation. These constraints required a uniformly observed quasar sample density of $>58$\,\degdeg\ for the redshift range $0.9 < z < 2.2$. A $z > 2.1$ quasar sample was also targeted to increase constraints on $D_A(z)$ and $H(z)$ in the \lya\ forest by $\sim1.44\times$ compared to SDSS-III/BOSS \citep{eboss_mission}. eBOSS achieved the targeted precision for the lower redshift range \citep{hou20a,neveux20a,smith20} and the \lya\ forest sample \citep{bourboux20a}. The targeting program to select the eBOSS quasar sample, which is detailed in \citet{eBOSS_target_select}, is summarized below.

The majority of eBOSS quasars were targeted by a {\tt CORE} algorithm, which was applied to SDSS {\tt SURVEY\_PRIMARY} point sources with (extinction-corrected) $g<22$ or $r<22$. These point sources were passed to the XDQSOz algorithm \citep{bovy2012}, which imposed a probability of being a quasar at redshifts of $z > 0.9$ of more than 20\%. An additional WISE-optical color cut was then applied to further reduce stellar contamination. These two selection criteria led to a quasar sample density of $\sim 70$\,\degdeg. Note that no explicit upper limit on redshift was applied, allowing the {\tt CORE} sample to also target \lya-forest quasars.

To constrain cosmological parameters, eBOSS targeted $z > 2.1$ quasars purely as back-lights of the \lya\ forest, meaning that such quasars could have a heterogeneous angular selection function. The {\tt CORE} sample was therefore supplemented by three diverse methods designed to increase the yield of \lya-forest quasars. Objects with $r > 19$ and $g < 22.5$ that displayed quasar-like variability in the PTF light curves discussed in \S\ref{ssec:imaging} were included as {\tt QSO\_PTF} targets, increasing the sample density to $\sim 74$\,\degdeg. The $<1$\,\degdeg\ of SDSS point sources that lie within 1\arcsec\ of a FIRST radio source were included as \texttt{QSO\_EBOSS\_FIRST} targets. Finally, \lya\ quasars that yielded a low S/N spectrum in BOSS were added as \texttt{QSO\_REOBS} targets, which increased the on-sky density by between $\sim 6$\,\degdeg\ and $\sim10$\,\degdeg.

\begin{figure}[ht]
    \epsscale{1.18}
    \plotone{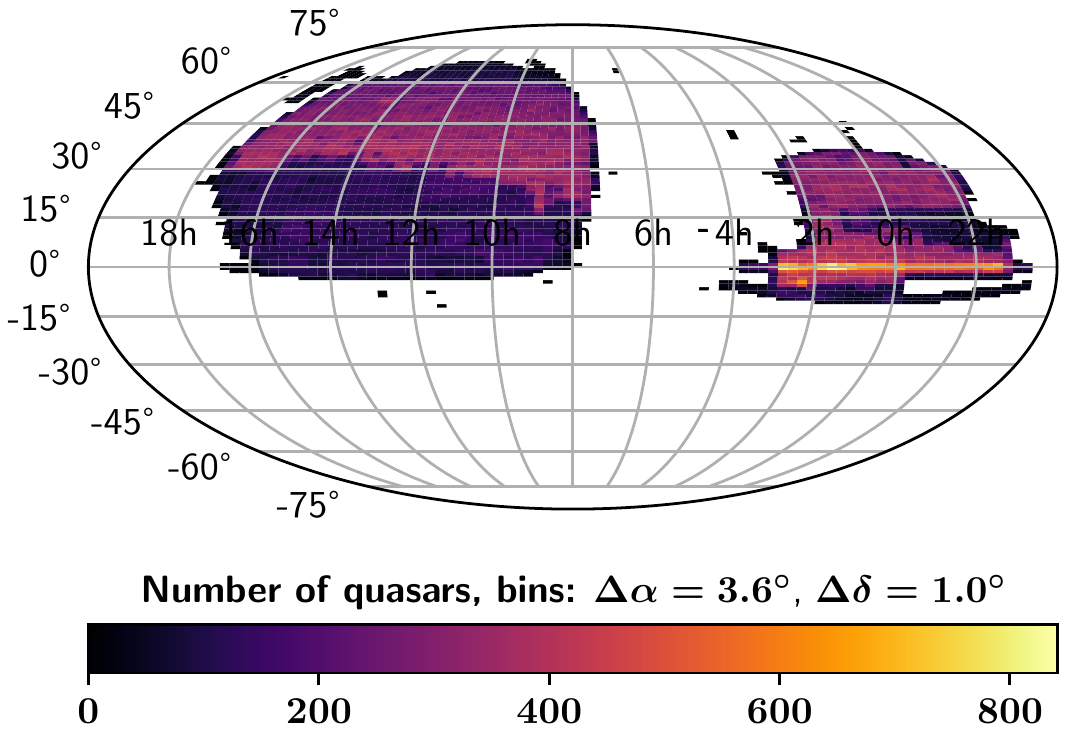}
    \caption{Quasars in DR16Q from SDSS-I/II, BOSS, and eBOSS. Right ascension ($\alpha$) and declination ($\delta$) are mapped onto a Mollweide projection of the sky.
    \label{fig:skymap}
    }
\end{figure}

Quasars that were targeted by the SDSS-IV sub-programs, TDSS \citep[detailed in][]{morganson2015, macleod2018} and SPIDERS \citep[detailed in][]{dwelly2017,comparat2019} are also included in DR16Q. The on-sky distribution of DR16Q quasars, which reflects all of the various targeting programs outlined in this sub-section, can be seen in Fig.~\ref{fig:skymap}.

\subsection{Spectroscopy}\label{ssec:spectroscopy}
Spectroscopy for SDSS-IV/eBOSS was conducted using the BOSS spectrographs
\citep{smee2013} on the 2.5\,m Sloan Telescope. Two spectrographs each recorded data from 500 fibers on a 2k CCD with square 24\,\micm\ pixels and a wavelength range of 3600\,\AA\ to 10{,}400\,\AA\ at a spectral resolution of $\lambda / \Delta \lambda \approx 2000$.

Each calibrated group of 1000 spectra were processed by the BOSS \texttt{spec1d} pipeline \citep[][see also \S\ref{ssec:automatedredshifts}]{bolton2012}. Spectra were fit using a variable number of rest-frame-derived PCA templates, which were applied using least-squares minimization to find the five best quasar redshifts, five best galaxy redshifts, 123 stellar redshifts, and one cataclysmic variable star redshift. The fits were then ranked according to the smallest reduced $\chi^{2}$ ($\chi_{r}^{2}$) value. The redshift, object classification, and line identifications were taken from the spectral fit with the lowest $\chi_{r}^{2}$. In the case that two spectral fits had a $\chi_{r}^{2}$ difference less than 0.01 a \texttt{ZWARNING} flag was assigned.

\begin{figure}[ht]
    \epsscale{1.18}
    \plotone{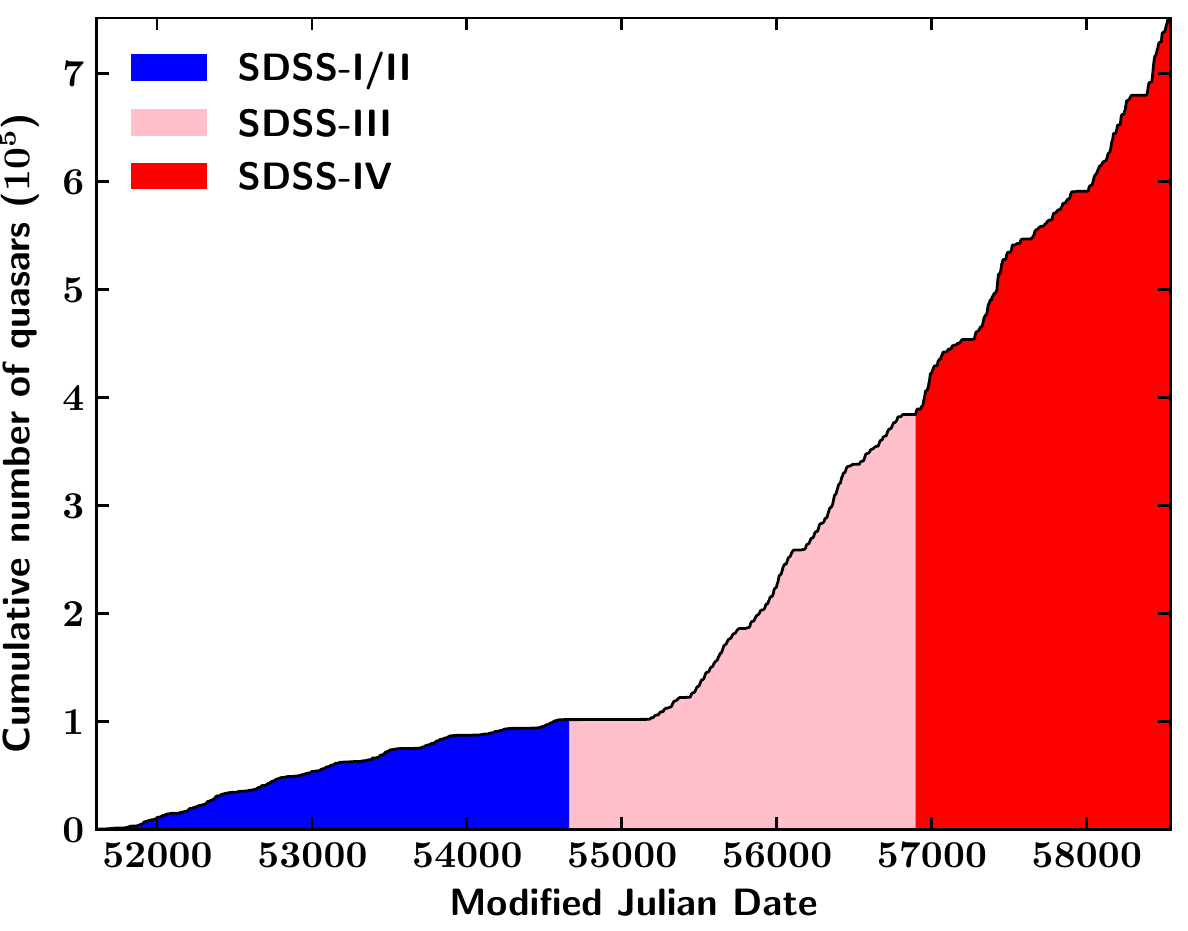}
    \caption{The cumulative number of quasars observed over the four SDSS campaigns. SDSS-I/II is shaded in blue ($\textrm{MJD}< 54663$), SDSS-III in pink ($54663 \leq \textrm{MJD} < 56898$), and SDSS-IV in red ($56898 \leq \textrm{MJD} < 58543$). Quasars observed multiple times were included based on their first spectroscopic observation. DR16Q includes 750{,}414 confirmed, quasars. Flat regions of the plot indicate periods of shutdown.
    \label{fig:barsum}}
\end{figure}

DR16Q includes spectra obtained using both the BOSS spectrographs and the  SDSS-I/II spectrographs. The cumulative number of observed quasar spectra by campaign is shown in Fig.~\ref{fig:barsum}. Wavelength limits for the spectra differ between the two set-ups, with SDSS-I/II spectra covering 3800\,\AA\ to 9100\,\AA\ \citep[see][]{smee2013}.

\section{Construction of the catalog}\label{sec:construct}
As DR16Q represents the final SDSS-IV quasar catalog, it contains all quasars observed as part of eBOSS, TDSS, and SPIDERS. Additionally, like the catalog of \citet[][henceforth DR14Q]{dr14q_paper}, quasars observed in SDSS-I/II, and SDSS-III/BOSS have been included. Quasars presented in \citet[][henceforth DR7Q]{dr7q_paper} and \citet[][henceforth DR12Q]{dr12q_paper} that did not have an eBOSS observation were added via coordinate-matching, as detailed in \S\ref{ssec:def_superset}. Due to quasar identification errors, quasars in DR14Q without a match in DR16Q were not included in DR16Q (see \S\ref{asec:oldbugs} for details). The {\tt python} code for generating DR16Q is available publicly.\footnote{\url{https://github.com/bradlyke/dr16q}}

\subsection{Definition of the Superset}\label{ssec:def_superset}
The DR16Q superset was constructed from \texttt{spAll-v5\_13\_0.fits} (henceforth ``the spAll file''), the file of {\em all} SDSS-III/IV observations generated by version \texttt{v5\_13\_0} of the SDSS spectroscopic pipeline\footnote{\url{https://data.sdss.org/datamodel/files/BOSS\_SPECTRO\_REDUX/RUN2D/spAll.html}}. Any observation in the spAll file that has a bit flagged that is recorded in the columns listed in Table~\ref{tab:tarbits} was included in the DR16Q superset.

As the spAll file does not include observations from SDSS-I/II, quasars from DR7Q without an eBOSS reobservation were added to the DR16Q superset. Additionally, some quasars appearing in DR12Q were serendipitous identifications and did not have the bits set from Table~\ref{tab:tarbits}. DR7Q and DR12Q were coordinate-matched with a $0.5\arcsec$~radius to the superset, and missing objects were added. If a quasar already appeared in the superset, only the redshift and spectroscopic identifiers (plate, MJD, and fiber ID) were added to the SDSS-III/IV record. Known astrometric errors in DR12Q were corrected before coordinates were matched. The DR7Q and DR12Q quasars were added to the superset {\em after} the winnowing algorithm outlined in \S\ref{ssec:autoclass} was applied, as any objects appearing in DR7Q and DR12Q had previously been visually confirmed to be confident quasars. For quasars included in both DR7Q and DR12Q, only the DR7Q observation was propagated to this catalog superset.

The superset contains 1{,}440{,}615 observations of quasars, stars, and galaxies that were all targeted as quasars (or appeared in previous quasar catalogs).

\renewcommand{\arraystretch}{0.9}
\begin{deluxetable*}{L >{\tt}l|L >{\tt}l|L >{\tt}l|L >{\tt}l}[hb!]
\tablecaption{Targeting Bit Parameters \label{tab:tarbits}}
\tablecolumns{8}
\tablewidth{0pt}
\tablehead{
\colhead{Bit} & \colhead{Selection} & \colhead{Bit} & \colhead{Selection} & \colhead{Bit} & \colhead{Selection} & \colhead{Bit} & \colhead{Selection}
}
\startdata
\hline
\hline
\multicolumn{8}{l}{BOSS\_TARGET1} \\
\hline
10 & QSO\_CORE$^{(1)}$ & 11 & QSO\_BONUS$^{(1)}$ & 12 & QSO\_KNOWN\_MIDZ$^{(1)}$ & 13 & QSO\_KNOWN\_LOHIZ$^{(1)}$ \\
14 & QSO\_NN$^{(1)}$ & 15 & QSO\_UKIDSS$^{(1)}$ & 16 & QSO\_KDE\_COADD$^{(1)}$ & 17 & QSO\_LIKE$^{(1)}$ \\
18 & QSO\_FIRST\_BOSS$^{(1)}$ & 19 & QSO\_KDE$^{(1)}$ & 40 & QSO\_CORE\_MAIN$^{(1)}$ & 41 & QSO\_BONUS\_MAIN$^{(1)}$ \\
42 & QSO\_CORE\_ED$^{(1)}$ & 43 & QSO\_CORE\_LIKE$^{(1)}$ & 44 & QSO\_KNOWN\_SUPPZ$^{(1)}$ & & \\
\hline
\multicolumn{8}{l}{EBOSS\_TARGET0} \\
\hline
10 & QSO\_EBOSS\_CORE$^{(5)}$ & 11 & QSO\_PTF$^{(5)}$ & 12 & QSO\_REOBS$^{(5)}$ & 13 & QSO\_EBOSS\_KDE$^{(5)}$ \\
14 & QSO\_EBOSS\_FIRST$^{(5)}$ & 15 & QSO\_BAD\_BOSS$^{(5)}$ & 16 & QSO\_BOSS\_TARGET$^{(5)}$ & 17 & QSO\_SDSS\_TARGET$^{(5)}$ \\
18 & QSO\_KNOWN$^{(5)}$ & 20 & SPIDERS\_RASS\_AGN$^{(6)}$ & 22 & SPIDERS\_ERASS\_AGN$^{(6)}$ & 30 & TDSS\_A$^{(7)}$ \\
31 & TDSS\_FES\_DE$^{(8)}$ & 33 & TDSS\_FES\_NQHISN$^{(8)}$ & 34 & TDSS\_FES\_MGII$^{(8)}$ & 35 & TDSS\_FES\_VARBAL$^{(8)}$ \\
40 & SEQUELS\_PTF\_VARIABLE & & & & & & \\
\hline
\multicolumn{8}{l}{EBOSS\_TARGET1} \\
\hline
9 & QSO1\_VAR\_S82$^{(9)}$ & 10 & QSO1\_EBOSS\_CORE$^{(5)}$ & 11 & QSO1\_PTF$^{(5)}$ & 12 & QSO1\_REOBS$^{(5)}$ \\
13 & QSO1\_EBOSS\_KDE$^{(5)}$ & 14 & QSO1\_EBOSS\_FIRST$^{(5)}$ & 15 & QSO1\_BAD\_BOSS$^{(5)}$ & 16 & QSO\_BOSS\_TARGET$^{(5)}$ \\
17 & QSO\_SDSS\_TARGET$^{(5)}$ & 18 & QSO\_KNOWN$^{(5)}$ & 30 & TDSS\_TARGET$^{(7,8,10)}$ & 31 & SPIDERS\_TARGET$^{(6,10)}$ \\
\hline
\multicolumn{8}{l}{EBOSS\_TARGET2} \\
\hline
0 & SPIDERS\_RASS\_AGN$^{(6)}$ & 2 & SPIDERS\_ERASS\_AGN$^{(6)}$ & 4 & SPIDERS\_XMMSL\_AGN$^{(6)}$ & 20 & TDSS\_A$^{(7)}$ \\
21 & TDSS\_FES\_DE$^{(8)}$ & 23 & TDSS\_FES\_NQHISN$^{(8)}$ & 24 & TDSS\_DES\_MGII$^{(8)}$ & 25 & TDSS\_FES\_VARBAL$^{(8)}$ \\
26 & TDSS\_B$^{(7)}$ & 27 & TDSS\_FES\_HYPQSO$^{(8)}$ & 31 & TDSS\_CP$^{(8)}$ & 32 & S82X\_TILE1$^{(10)}$ \\
33 & S82X\_TILE2$^{(10)}$ & 34 & S82X\_TILE3$^{(10)}$ & 50 & S82X\_BRIGHT\_TARGET$^{(10)}$ & 51 & S82X\_XMM\_TARGET$^{(10)}$ \\
52 & \multicolumn{3}{l|}{\tt{S82X\_WISE\_TARGET}$^{(10)}$}  & 53 & \multicolumn{3}{l}{\tt{S82X\_SACLAY\_VAR\_TARGET}$^{(10)}$} \\
54 & \multicolumn{3}{l|}{\tt{S82X\_SACLAY\_BDT\_TARGET}$^{(10)}$} & 55 & \multicolumn{3}{l}{\tt{S82X\_SACLAY\_HIZ\_TARGET}$^{(10)}$} \\
56 & \multicolumn{3}{l|}{\tt{S82X\_RICHARDS15\_PHOTOQSO\_TARGET}$^{(10)}$} & 57 & \multicolumn{3}{l}{\tt{S82X\_PETERS15\_COLORVAR\_TARGET}$^{(10)}$} \\
58 & \multicolumn{3}{l|}{\tt{S82X\_LSSTZ4\_TARGET}$^{(10)}$}  & 59 & \multicolumn{3}{l}{\tt{S82X\_UNWISE\_TARGET}$^{(10)}$} \\
60 & \multicolumn{3}{l|}{\tt{S82X\_GTRADMZ4\_TARGET}$^{(10)}$}  & 61 & \multicolumn{3}{l}{\tt{S82X\_CLAGN1\_TARGET}$^{(10)}$} \\
62 & \multicolumn{3}{l|}{\tt{S82X\_CLAGN2\_TARGET$^{(10)}$}}  & & \multicolumn{3}{l}{} \\
\hline
\multicolumn{8}{l}{ANCILLARY\_TARGET1} \\
\hline
6 & BLAZGVAR$^{(2)}$ & 7 & BLAZR$^{(2)}$ & 8 & BLAZXR$^{(2)}$ & 9 & BLAZXRSAL$^{(2)}$ \\
10 & BLAZXRVAR$^{(2)}$ & 11 & XMMBRIGHT$^{(2)}$ & 12 & XMMGRIZ$^{(2)}$ & 13 & XMMHR$^{(2)}$ \\
14 & XMMRED$^{(2)}$ & 15 & FBQSBAL$^{(2)}$ & 16 & LBQSBAL$^{(2)}$ & 17 & ODDBAL$^{(2)}$ \\
18 & OTBAL$^{(2)}$ & 19 & PREVBAL$^{(2)}$ & 20 & VARBAL$^{(2)}$ & 22 & QSO\_AAL$^{(2)}$ \\
23 & QSO\_AALS$^{(2)}$ & 24 & QSO\_IAL$^{(2)}$ & 25 & QSO\_RADIO$^{(2)}$ & 26 & QSO\_RADIO\_AAL$^{(2)}$ \\
27 & QSO\_RADIO\_IAL$^{(2)}$ & 28 & QSO\_NOAALS$^{(2)}$ & 29 & QSO\_GRI$^{(2)}$ & 30 & QSO\_HIZ$^{(2)}$ \\
31 & QSO\_RIZ$^{(2)}$ & 50 & BLAZGRFLAT$^{(2)}$ & 51 & BLAZGRQSO$^{(2)}$ & 52 & BLAZGX$^{(2)}$ \\
53 & BLAZGXQSO$^{(2)}$ & 54 & BLAZGXR$^{(2)}$ & 55 & BLAZXR$^{(2)}$ & 58 & CXOBRIGHT$^{(2)}$ \\
59 & CXORED$^{(2)}$ & & & & & & \\
\hline
\multicolumn{8}{l}{ANCILLARY\_TARGET2} \\
\hline
0 & HIZQSO82$^{(2)}$ & 1 & HIZQSOIR$^{(2)}$ & 2 & KQSO\_BOSS$^{(2)}$ & 3 & QSO\_VAR$^{(2)}$ \\
4 & QSO\_VAR\_FPG$^{(2)}$ & 5 & RADIO\_2LOBE\_QSO$^{(2)}$ & 7 & QSO\_SUPPZ$^{(2)}$ & 8 & QSO\_VAR\_SDSS$^{(2)}$ \\
9 & QSO\_WISE\_SUPP$^{(3)}$ & 10 & QSO\_WISE\_FULL\_SKY$^{(4)}$ & 13 & DISKEMITTER\_REPEAT$^{(4)}$ & 14 & WISE\_BOSS\_QSO$^{(4)}$ \\
15 & QSO\_XD\_KDE\_PAIR$^{(4)}$ & 24 & TDSS\_PILOT$^{(4)}$ & 25 & SPIDERS\_PILOT$^{(4)}$ & 26 & TDSS\_SPIDERS\_PILOT$^{(4)}$ \\
27 & QSO\_VAR\_LF$^{(4)}$ & 31 & QSO\_EBOSS\_W3\_ADM$^{(4)}$ & 32 & XMM\_PRIME$^{(4)}$ & 33 & XMM\_SECONDS$^{(4)}$ \\
53 & SEQUELS\_TARGET$^{(4)}$ & 54 & RM\_TILE1$^{(4)}$ & 55 & RM\_TILE2$^{(4)}$ & 56 & QSO\_DEEP$^{(4)}$ \\
\hline
\multicolumn{8}{l}{$^{(1)}$\citet{boss_target_select} -- $^{(2)}$\citet{boss_mission} -- $^{(3)}$\citet{ancil_target2_suppflag} -- $^{(4)}$\citet{ancil_target2_flags} -- $^{(5)}$\citet{eBOSS_target_select}}\\
\multicolumn{8}{l}{$^{(6)}$\citet{dwelly2017} -- $^{(7)}$\citet{morganson2015} -- $^{(8)}$\citet{macleod2018} -- $^{(9)}$\citet{palanque2016}}\\
\multicolumn{8}{l}{$^{(10)}$\citet{abolfathi2018}}\\
\enddata
\end{deluxetable*}

\subsection{Automated classification}\label{ssec:autoclass}
Spectra flagged as in Table~\ref{tab:tarbits} were selected from the spAll file and passed through a modified version of a classification scheme that first appeared in \citet{eboss_mission}. The ``top 5'' classifications (those with the 5 lowest $\chi_{r}^{2}$; see \S\ref{ssec:spectroscopy}) are used in the following decision tree (in order) to set the value of the \texttt{AUTOCLASS\_DR14Q} field:
\begin{enumerate}
    \item If the best model fit for the spectrum is \texttt{STAR}, the spectrum is classified as a \texttt{STAR}.
    \item If the best model fit for the spectrum is \texttt{GALAXY} and \zpipe~$< 1$, then the spectrum is classified as a \texttt{GALAXY}.
    \item If the best model fit for the spectrum is \texttt{GALAXY}, \zpipe~$\geq 1$, and at least one other fit is \texttt{GALAXY}, then the spectrum is classified as a \texttt{GALAXY}.
    \item If the best model fit for the spectrum is \texttt{QSO} and 2 or more other model fits are \texttt{STAR}, then the spectrum is classified as a \texttt{STAR}.
    \item If the best model fit for the spectrum is \texttt{QSO}, fewer than 2 other model fits are \texttt{STAR}, and $\texttt{ZWARNING} = 0$, the spectrum is classified as a \texttt{QSO}.
    \item If a spectrum meets none of these criteria, it is selected for visual inspection (\texttt{AUTOCLASS\_DR14Q} = \texttt{VI}).
\end{enumerate}

Each individual spectrum was classified using this schema, and, afterwards, any object that had an initial classification of \texttt{QSO} and \zpipe~$>3.5$ was reclassified for visual inspection (\texttt{AUTOCLASS\_DR14Q} = \texttt{VI}). This algorithm flagged $\sim 6\%$ of the superset ($\sim 87{,}000$ spectra) for visual inspection. To reduce duplication of effort, we only visually inspected spectra taken after MJD 57905, the last date of inspections for DR14Q. This reduced our flagged percentage to $\sim 1.5\%$, or 20{,}508 spectra.

To further reduce the number of visual inspections, the \qnet\ algorithm \citep{busca2018} was applied. \qnet\ produces a binary quasar flag and a redshift, which we designated \texttt{IS\_QSO\_QN} and \texttt{Z\_QN} respectively. We used the results stored in \texttt{AUTOCLASS\_DR14Q} to reclassify objects based on confident \qnet\ classifications, recording the output in \texttt{AUTOCLASS\_PQN}. \texttt{AUTOCLASS\_PQN} retained the same classification as \texttt{AUTOCLASS\_DR14Q} for all cases but one: if \texttt{AUTOCLASS\_DR14Q} was \texttt{VI}, \texttt{IS\_QSO\_QN} $= 1$, and \texttt{Z\_QN} $<2.0$, then \texttt{AUTOCLASS\_PQN} was changed to \texttt{QSO}. Applying \qnet\ led to a reduction in the number of spectra to inspect to $\sim 0.6\%$ (8{,}581 spectra). Finally, objects were removed from the superset where the \texttt{ZWARNING}\footnote{\url{www.sdss.org/dr16/algorithms/bitmasks/\#ZWARNING}} field had one or more of the following flags set: \texttt{UNPLUGGED}, \texttt{SKY}, \texttt{LITTLE\_COVERAGE}, \texttt{BAD\_TARGET}, or \texttt{NODATA}.

To characterize the accuracy of our classification scheme, the SDSS-III/SEQUELS data set \citep[see \S5.1 of][]{eBOSS_target_select} was used as ``truth'' to calculate the sample completeness via,

\begin{equation}\label{eqn:completeness}
    {\rm{Comp}} = \frac{N_{\rm{pq}} + N_{\rm{vq}}}{N_{\rm{tq}}},
\end{equation}
where $N_{\rm{pq}}$ is the number of quasars correctly classified as \texttt{QSO} by our schema, $N_{\rm{vq}}$ is the number of quasars identified during visual inspections, and $N_{\rm{tq}}$ is the total number of quasars in the catalog. We also calculated the sample contamination via,
\begin{equation}\label{eqn:contamination}
    {\rm{Contam}} = \frac{N_{\rm{nq}}}{N_{\rm{tq}} + N_{\rm{nq}}},
\end{equation}
where $N_{\rm{nq}}$ is the number of spectra incorrectly classified as quasar spectra and $N_{\rm{tq}}$ is as in Eqn.~\ref{eqn:completeness}. We calculated a completeness of $99.2\%$ and a contamination of $0.5\%$. Due to increased observation time and subsequent higher signal-to-noise ratios for these spectra, the BOSS SEQUELS set is not necessarily a good indicator of the spectral quality throughout eBOSS.

\subsection{Random eBOSS Visual Inspections}\label{ssec:randomvi}
To better characterize the eBOSS spectral pipeline, a random subsample of 10{,}000 eBOSS spectra from the DR16Q superset was selected. This subsample consisted of objects from the eBOSS {\tt CORE} set and non-{\tt CORE} objects with a pipeline redshift greater than 1.8. This subsample was visually inspected to check whether the pipeline correctly classified the spectrum and, in the case of quasar spectra, assigned an accurate redshift (``accurate'' being defined as $\Delta v \leq 3000$\,\kms; see also \S\ref{ssec:redshiftcomparisons}). Whether the pipeline's classification (and redshift) was correct is recorded in DR16Q in the column \texttt{PIPE\_CORR\_10K}. If the pipeline was correct, \texttt{Z\_10K} was set to the pipeline redshift, and the pipeline's determination of a quasar classification was retained.

If the pipeline did not correctly classify the spectrum, or a quasar redshift was significantly wrong ($\Delta v > 3000$\,\kms), the classification was corrected and quasar redshifts were recalculated as detailed in \S\ref{ssec:eboss_vi}. These data are recorded in DR16Q in the columns \texttt{IS\_QSO\_10K} for classification, \texttt{Z\_10K} for the redshift, and \texttt{Z\_CONF\_10K} for the redshift confidence in the case of quasars. Eqn.~\ref{eqn:completeness} was applied to this set, producing a completeness of $99.8\%$. Some of the random subsample of 10{,}000 eBOSS spectra with a low signal-to-noise ratio were classified as quasars by the DR14Q visual inspection team but were reclassified as non-quasars by the DR16Q team. Applying Eqn.~\ref{eqn:contamination} and using DR14Q classifications for such cases produced an estimated contamination of $0.3\%$. Using DR16Q classifications for such cases produced an estimated contamination of $1.3\%$.

\subsection{Visual inspection in eBOSS}\label{ssec:eboss_vi}
In past SDSS visually inspected catalogs, such as DR7Q \citep{dr7q_paper} and DR12Q \citep{dr12q_paper}, the number of quasar candidates was small enough for each spectrum to be visually inspected. In eBOSS the number of observations rose precipitously, so, for DR16Q, we only visually inspected objects as outlined in \S\ref{ssec:autoclass}. As in DR14Q, objects selected for visual inspection were reclassified and quasar redshifts were corrected where needed. Spectra that were not quasars upon visual inspection, or that had a very low signal-to-noise ratio, were classified as ``non-quasars'' and did not have their redshifts corrected. These appear in the superset with \texttt{CLASS\_PERSON} $=4$ (\texttt{GALAXY}), and have $\texttt{Z\_VI} = -1$. The numeric system used to classify spectra in DR16Q is shown in Table~\ref{tab:classperson}. This system is similar to that used in DR12Q, but the (unused) value of 2 was removed for DR16Q and a new value of 50 was added to indicate potential blazars.

\begin{deluxetable}{Cl | Cl}[ht]
\tablecaption{Visual inspection classifications, which appear in the \texttt{CLASS\_PERSON} column of DR16Q. \label{tab:classperson}}
\tablecolumns{4}
\tablewidth{0pt}
\tablehead{
\colhead{Value} & \colhead{Object Type} & \colhead{Value} & \colhead{Object Type}
}
\startdata
0 & Not Inspected & 1 & Star \\
3 & Quasar & 4 & Galaxy \\
30 & BAL Quasar & 50 & Blazar(?)\\
\enddata
\end{deluxetable} \vspace{-1cm}

Any spectra flagged for visual inspection in DR16Q were classified as a ``non-quasar'' if the noise spectrum was greater than the continuum everywhere and no more than one emission line peak rose above the noise level. If two emission line peaks could be positively identified, the object was classified as a quasar and given a corrected, ``visual inspection'' redshift. In a few cases, a single evident emission line peak could be questionably identified. Where this was true, the redshift was corrected using that emission line, the object was classified as a quasar, but the confidence rating for the redshift (\texttt{Z\_CONF}; described below) was set to 1. During visual inspection, we identified four spectra with high flux, continua matching an archetypal quasar continuum shape, and no identifiable absorption or emission features. These spectra were marked as possible blazars; they were assigned a \texttt{CLASS\_PERSON} value of 50, and a redshift of -999.

Where emission line peaks (or other features) could be identified, a new redshift was calculated by using the peak (or feature) wavelength for the following emission lines in priority order: \mgii\ $\lambda2799$, \civ\ $\lambda1549$, \ciii\ $\lambda1908$, \oiii\ $\lambda5007$, \oii\ $\lambda3728$, and Lyman break $\lambda912$. A confidence rating for these redshifts was recorded in the column \texttt{Z\_CONF}, with 0 being the lowest and 3 being the highest confidence (a value of -1 indicates the object was not visually inspected). Spectra with one questionable emission line were given a confidence rating of 1. Spectra with two or more lines identified based on pipeline locations were given a confidence rating of 2. In all cases, visual inspection redshifts were calculated to three decimal places. If the visual inspection redshift agreed with the pipeline, the visual inspection redshift was set to the pipeline redshift.

Unlike for DR12Q, spectra of general interest such as those of damped Lyman-$\alpha$ systems (DLAs) and broad absorption line quasars (BALs) were not flagged as part of the visual inspection process. Instead, DLAs and BALs were identified via algorithm, as described in \S\ref{sec:dlabal}.

As the SDSS classification pipeline changed substantially between DR14Q and DR16Q, DR16Q only includes visual inspection information from DR14Q for objects that were in the DR16Q superset. Additionally, objects in DR16Q with $\texttt{Z} > 5$ and $\texttt{SOURCE\_Z} = \rm{PIPE}$ should be considered suspect, as they sometimes have misleading spectroscopic reductions or classifications. Users should carefully reinspect such objects prior to using them for scientific analyses.

\subsection{Duplicate Observation Removal}\label{ssec:duplicateremove}
Some SDSS-IV/eBOSS quasar targets were part of multi-epoch campaigns (e.g., Reverberation Mapping;~\citealt{shen2015}; TDSS;~\citealt{morganson2015, macleod2018}). Additionally, many targets observed in previous SDSS campaigns were reobserved in eBOSS. To reduce the superset of observations to {\em objects} we employed the following algorithm:

\begin{enumerate}
    \item The superset catalog was coordinate-matched to itself with a maximum matching radius of $0.5\arcsec$ and self-matches were removed.
    \item For objects that had at least one visual inspection, we selected the observation with the highest VI confidence rating.
    \item If all of the confidence ratings were the same, or none of the observations had a visual inspection, we selected the observation with the highest \texttt{SN\_MEDIAN\_ALL} value.
\end{enumerate}

The \texttt{SN\_MEDIAN\_ALL} value represents the median signal-to-noise ratio across all good pixels in a spectrum. Primary observations were marked with a 1 in the \texttt{PRIM\_REC} column of the DR16Q superset. For non-primary observations the spectroscopic plate, modified Julian date, fiber ID number, and spectroscopic instrument (SDSS or BOSS) were recorded in the fields of the primary observation marked \texttt{PLATE\_DUPLICATE}, \texttt{MJD\_DUPLICATE}, \texttt{FIBERID\_DUPLICATE}, and \texttt{SPECTRO\_DUPLICATE} respectively. The number of duplicate observations for each spectroscopic set-up were also tallied and recorded in \texttt{NSPEC\_SDSS} and \texttt{NSPEC\_BOSS}. The column \texttt{NSPEC} records the total number of duplicate observations. All duplicate observations are included as individual records in the DR16Q superset but only quasars with \texttt{PRIM\_REC} set to 1 were used to create the DR16Q quasar-only catalog.

\subsection{Classification results}\label{ssec:classresults}
DR16Q primarily contains data from eBOSS, but quasars appearing in DR7Q and DR12Q were also added (to both the DR16Q superset and the quasar-only catalog) if they did not already appear in the DR16Q superset. Due to the varied sources for the superset records, a more robust final classification for quasars was needed. A new column titled \texttt{IS\_QSO\_FINAL} was created that can take integer values from -2 to 2. In brief, quasar spectra have a value of 1, and questionable quasar spectra a value of 2. All values of 0 or less denote non-quasars. We developed a new algorithm to merge all of the independent, confident quasar classification sources into a final value. This final classification was generated by the following algorithm after the removal of duplicate observations:
\begin{enumerate}
    \item If \texttt{AUTOCLASS\_PQN} was \texttt{QSO}, the object is a quasar and \texttt{IS\_QSO\_FINAL} $= 1$.
    \begin{enumerate}
        \item However, if \texttt{CLASS\_PERSON} = 1 or 4 and \texttt{Z\_CONF} $\geq 2$, the object was not a quasar and \texttt{IS\_QSO\_FINAL} $= -2$.
    \end{enumerate}
    \item If \texttt{Z\_VI} or \texttt{Z\_10K} $= -999$, these are possible blazars and \texttt{IS\_QSO\_FINAL} $= 1$.
    \item If \texttt{AUTOCLASS\_PQN} = \texttt{UNK}, \texttt{CLASS\_PERSON} = 3 or 30, and \texttt{Z\_CONF} $\geq 2$, the object is a quasar and \texttt{IS\_QSO\_FINAL} $= 1$.
    \begin{enumerate}
        \item If the above is true, but \texttt{Z\_CONF} $= 1$, then the classification as a quasar is questionable, but included. We set \texttt{IS\_QSO\_FINAL} $= 2$. There were 69 objects in this subset.
    \end{enumerate}
    \item If \texttt{SOURCE\_Z} is \texttt{DR12QV}, \texttt{DR7QV\_SCH}, or \texttt{DR6Q\_HW}, the object was visually confirmed to be a quasar in a previous catalog and we set \texttt{IS\_QSO\_FINAL} $= 1$.
    \item If \texttt{RANDOM\_SELECT} $= 1$ and \texttt{IS\_QSO\_10K} $= 1$, the object is visually confirmed to be a quasar and we set \texttt{IS\_QSO\_FINAL} $= 1$.
    \item If \texttt{IS\_QSO\_DR12Q} or \texttt{IS\_QSO\_DR7Q} $= 1$, we set \texttt{IS\_QSO\_FINAL} $= 1$. This subset occurred when an object had an eBOSS primary observation and a duplicate DR12Q or DR7Q observation.
    \item If \texttt{AUTOCLASS\_PQN} = \texttt{VI}, \texttt{CLASS\_PERSON} = 3, 30, or 50, and \texttt{Z\_CONF} $\geq 2$, then we set \texttt{IS\_QSO\_FINAL} $= 1$.
    \begin{enumerate}
        \item However, if this last condition is true except \texttt{Z\_CONF} $= 1$, the object is questionably a quasar and we set \texttt{IS\_QSO\_FINAL} $= 2$. There were 670 objects in this subset.
    \end{enumerate}
\end{enumerate}
\texttt{AUTOCLASS\_PQN}, \texttt{CLASS\_PERSON}, \texttt{Z\_VI}, \texttt{Z\_10K}, \texttt{SOURCE\_Z}, \texttt{RANDOM\_SELECT}, and the various \texttt{IS\_QSO\_YYY} columns are described in detail in \S\ref{sec:description}. Once the \texttt{IS\_QSO\_FINAL} column was populated in the DR16Q superset, only objects with $\texttt{IS\_QSO\_FINAL} > 0$ were selected for the quasar-only catalog.

Two special cases existed after assigning classifications: First, an object classified as a quasar by the algorithm in \S\ref{ssec:autoclass} that had a low-confidence visual inspection classifying it as a non-quasar. Such objects were left in the quasar-only catalog but are possible contaminants. Second, an object classified as a quasar by a low-confidence visual inspection that was classified as a non-quasar by the automated algorithm. Such objects were removed and represent possible lost quasars. In other words, confident visual inspection classifications {\em would} override automated classifications, but low-confidence visual classifications would {\em not}.

Table~\ref{tab:classresults} lists the number of observations and objects appearing in both the DR16Q superset and the DR16Q quasar-only catalog.

\begin{deluxetable}{l c}[ht!]
\tablecaption{Classification Results \label{tab:classresults}}
\tablecolumns{2}
\tablewidth{0pt}
\tablehead{\multicolumn{2}{l}{\bf{Superset Numbers}}}
\startdata
Observations from SDSS-I/II & 73{,}325 \\
Observations from SDSS-III/IV & 1{,}367{,}290 \\
\textbf{Total observations in Superset} & \textbf{1{,}440{,}615} \\
\hline
\hline
Quasar observations in Superset & 920{,}110 \\
Duplicate quasar observations in DR16Q & 199{,}904 \\
\hline
\hline
\multicolumn{2}{l}{\bf{Quasar-only Numbers}}\\
\hline
Quasars with automated redshift only & 341{,}622 \\
Quasars with a visual inspection redshift & 408{,}792\\
\textbf{Total DR16Q quasars} & \textbf{750{,}414} \\
\enddata
\end{deluxetable}

\section{Redshift estimates}\label{sec:redshifts}
A homogeneously defined set of quasar redshifts is integral to the eBOSS mission of characterizing large-scale structure. Alternate redshift estimates, however, may be more useful for other science needs. We have therefore chosen to include a number of different redshift estimates in DR16Q, which we detail in this section.

\subsection{Automated redshifts}\label{ssec:automatedredshifts}
DR16Q includes automated classifications and redshifts determined by version \texttt{v5\_13\_0} of the SDSS spectroscopic pipeline. Observed spectra are fit by a set of models and templates that are detailed in \citet{bolton2012}. Quasar models are first fit by searching over redshifts binned in redshift space where bins are separated by four SDSS pixels in constant log-lambda spacing. The five models with the lowest $\chi^{2}$ are then fit to every pixel. These ``top five'' quasar fits are then compared, using reduced $\chi^{2}$ values, to the best five model fits for galaxies, 123 fits to stellar templates, and a fit to a cataclysmic variable template. Dubious pixels in spectra can be masked before fitting, but good spectra typically retain $\sim 4500$ pixels to fit. Quasar models include four eigenspectra components derived from BOSS reobservations of 568 SDSS DR5 quasars and a quadratic polynomial. In the version of the pipeline used to construct DR16Q, quasar models were optimized for BOSS, which targeted \lya\ forest quasars at about $z \geq 2.2$. In particular, the pipeline has trouble confidently distinguishing redshifts in the range $1.0 \leq z \leq 2.0$ for which the strong \oiii\ and \lya\ emission lines are not present in the eBOSS spectrum.

To check the quality of the pipeline classifications and redshifts, we compared the visual inspection redshifts derived from our random set of 10{,}000 superset spectra (\texttt{Z\_10K}; see \S\ref{ssec:randomvi}) to their pipeline values (\texttt{Z\_PIPE}) for eBOSS {\tt CORE} targets. Using a value of $\Delta v > 3000$\,\kms\ to define an inaccurate pipeline redshift (see also \S\ref{ssec:redshiftcomparisons}), we found that 2.1\% of the pipeline redshifts were inaccurate. This represented 154 catastrophic failures out of 7254 quasars. Further, the vast majority (130) of these catastrophes were highly inaccurate, where $\Delta v \geq 10{,}000$\,\kms.

\subsection{\qnet\ redshifts}
As discussed in \S\ref{ssec:autoclass}, we used redshifts from the \qnet\footnote{\url{https://github.com/ngbusca/QuasarNET}} algorithm
to help determine which quasars to visually inspect. We record this redshift in DR16Q as \texttt{Z\_QN}. For more information on \qnet, we refer the reader to \citet{busca2018}.

\subsection{Visual inspection redshifts}\label{ssec:viredshifts}
DR16Q includes the visually inspected redshifts for quasars that appeared in DR7Q or DR12Q in the \texttt{Z\_DR7Q\_SCH} and \texttt{Z\_DR12Q} columns. We also include the redshifts from \citet{dr6q_hw_paper}, where available, in the field \texttt{Z\_DR6Q\_HW}, as they are formally in the Hewett and Wild paper and have been used in some of our companion papers. We are conscious that a large body of work uses the updated DR7Q Hewett and Wild redshifts included in the ancillary columns of the value-added catalog detailed in \citet{shen2011}. We include these redshifts in their own column, \texttt{Z\_DR7Q\_HW}, for completeness. To conform with other reported redshifts, we do not include the \citet{shen2011} redshift errors in DR16Q.\footnote{See columns 143 and 144 at \url{http://das.sdss.org/va/qso_properties_dr7/dr7.htm}} For objects inspected after DR12Q, the visual inspections are included in the field \texttt{Z\_VI}. Redshifts for quasars found during the random visual inspection (see \S\ref{ssec:randomvi}), are recorded in the column \texttt{Z\_10K}. While some of these values may overlap we include each separately, as data reduction techniques may have changed between observations. In all cases, the field \texttt{SOURCE\_Z} records the origin of the primary redshift estimate, and the estimate itself is recorded in \texttt{Z} (as detailed in \S\ref{ssec:bestredshift}).

\subsection{PCA redshift and emission line redshifts}\label{ssec:systemvsemission}
In the tradition of previous BOSS and eBOSS quasar catalogs, DR16Q includes a redshift generated by principal component analysis (PCA), using the \rvb\ algorithm\footnote{\url{https://github.com/londumas/redvsblue}}. These redshifts are recorded in the \texttt{Z\_PCA} field of DR16Q.

In fitting the spectra, the \rvb\ algorithm uses the same four PCA eigenvectors as the SDSS spectroscopic pipeline and includes a second degree polynomial as a broadband term. Unlike the SDSS pipeline, \rvb\ uses all pixels within the observed wavelength range 3600\,\AA\ to 10{,}000\,\AA, regardless of whether they would be masked by the pipeline. The \rvb\ algorithm is also not limited to the pipeline redshift range of $0 \leq$ \zpipe\ $\leq 7$.

For quasars in the DR16Q quasar-only sample, we stacked all spectra that did not have the \texttt{ZWARNING}\footnote{\url{www.sdss.org/dr16/algorithms/bitmasks/\#ZWARNING}} flags \texttt{SKY}, \texttt{LITTLE\_COVERAGE}, \texttt{UNPLUGGED}, \texttt{BAD\_TARGET} or \texttt{NODATA} set, before assigning a PCA redshift. Spectra are 'stacked' by matching them in log-lambda space and taking an error-weighted average of the flux density at each point. For data appearing in the DR16Q superset, only the best observation was used. For both the DR16Q quasar-only catalog and the superset, the \rvb\ algorithm corrected all observed spectra for Galactic extinction using the dust map of \citet{schlegel1998}. The algorithm also corrected the model PCA eigenvectors for \lya\ transmission evolution using parameters from \citet{calura2012}. We used the primary DR16Q redshift (\texttt{Z}), with a flat prior of $\Delta v \pm 10{,}000$\,\kms, to seed an initial redshift estimate. In addition to the full-spectrum PCA redshift (\texttt{Z\_PCA}), we independently computed six emission line redshifts (\ha, \hb, \mgii, \ciii, \civ, and \lya) using the same prior on the primary redshift (\texttt{Z}) as used to calculate \texttt{Z\_PCA} but not using \texttt{Z\_PCA} itself\footnote{We use ``PCA redshift'' to denote a redshift not tied to a single emission line. Redshifts derived from a single emission line are referred to as ``emission line redshifts.''}. We include these redshifts in DR16Q when the emission line of interest is in the observed frame of the spectrum. A distinct PCA redshift was used in the \lya\ forest clustering catalogs. This redshift, recorded in the column \texttt{Z\_LYAWG} masks the \lya\ emission line and potential forest before generating a redshift. More details about this redshift can be found in \citet{bourboux20a}.

Some of the 750{,}414 sources in the DR16Q quasar-only sample did not yield good PCA redshifts. These included: 12{,}412 quasars where \texttt{Z\_PIPE} disagreed with \texttt{Z\_QN} by more than 10{,}000\,\kms; 1085 quasars that did not have an SDSS identifier ($\texttt{THING\_ID}\footnote{\url{https://www.sdss.org/dr16/algorithms/resolve/}} = -1$), so could not be stacked for the quasar-only catalog; 4 quasars with $\texttt{Z} = -999$ (possible blazars); and 444 that had outlier spectra that \rvb\ could not fit reliably. In addition, there are 11 quasars that have ${\tt Z\_PCA} < 0$. In total, 665{,}612 quasars in the quasar-only catalog yielded reliable PCA redshifts. The DR16Q superset contains 920{,}109 quasars with $\texttt{Z\_PCA} > 0$. These PCA redshifts represent a sample with homogeneous statistical and systematic errors.

\subsection{Selection of the ``Primary'' Redshift}\label{ssec:bestredshift}
DR16Q includes many different quasar redshift estimates. We select a ``primary'' redshift (similar to the ``best'' redshift in DR14Q) for each object from, most preferably, the available visual inspection redshifts, or, alternatively, the SDSS automated pipeline redshift. The columns \texttt{Z} and \texttt{SOURCE\_Z} record this primary redshift and from which column it was selected. See \S\ref{ssec:viredshifts} for more information on the available visual inspection redshifts.

For objects that have a redshift in the columns \texttt{Z\_VI} or \texttt{Z\_10K} and a confidence (\texttt{Z\_CONF} or \texttt{Z\_CONF\_10K}) of  $\geq2$, \texttt{Z} records the corresponding redshift and \texttt{SOURCE\_Z} is set to \text{VI}. Otherwise, if an object has a redshift in the columns \texttt{Z\_DR6Q\_HW} or \texttt{Z\_DR7Q\_SCH} these values are used (with \texttt{Z\_DR6Q\_HW} overriding \texttt{Z\_DR7Q\_SCH}) and \texttt{SOURCE\_Z} is set to \texttt{DR6Q\_HW} or \texttt{DR7QV\_SCH}. As the \texttt{Z\_DR7Q\_HW} redshifts did not formally appear in the \citet{shen2011} paper, these values are {\em not} used to populate the \texttt{Z} column. If no other visual inspection redshift is populated then \texttt{Z\_DR12Q} is used (and \texttt{SOURCE\_Z} is set to \texttt{DR12QV}). For objects with DR12Q redshifts, only the visual inspection redshifts are recorded; DR12Q pipeline redshifts are not included. In the absence of any of these visual inspection redshifts, \texttt{Z} is populated with the automated pipeline redshift (and \texttt{SOURCE\_Z} is set to \texttt{PIPE}).

The PCA and \qnet\ redshifts are included in their own columns in DR16Q but were not used to inform the \texttt{Z} column. Given the heterogeneous source information that is propagated into the \texttt{Z} column, we expect \texttt{Z} to represent the least biased redshift estimator, but with a high variance. For analyses that require a homogeneous redshift over a large ensemble we recommend \texttt{Z\_PCA}. We ourselves use \texttt{Z\_PCA} in this paper as a redshift prior for calculating absolute \textit{i}-band magnitudes, and for finding DLAs and BALs (\S\ref{sec:dlabal}).

\subsection{Comparison of redshift values}\label{ssec:redshiftcomparisons}
The number and complexity of physical processes that can affect the spectrum of a quasar make it difficult to precisely and accurately disentangle a ``systemic'' redshift (i.e., as a meaningful indicator of distance) from measured redshifts. Indeed, quasar spectra contain broad emission lines due to the rotating gas located around the central black hole that are subject to matter outflows around the accretion disk. These astrophysical processes frequently give rise to systematic offsets when measuring redshifts. To investigate the systematic shifts associated with the automated and emission line redshifts (\zpipe, \zpca, and $Z_{\textrm{\footnotesize{line}}}$), we used the Reverberation Mapping program \citep{shen2015} to compare the eBOSS redshifts to these ``systemic'' redshifts obtained from coadded-spectra of 849 confirmed quasars where $0.1 < z < 4.5$. \citet{shen2016} measured the velocity shifts of quasar emission line peaks compared to stellar absorption lines from the host galaxy of individual quasars. These stellar features correspond to absorption in the host galaxy which is not affected by contaminating physical processes providing reliable ``systemic'' velocity measurements. Thus the ``systemic'' redshifts are henceforth referred to as ``host'' redshifts. We discarded the first year of observations, to match the eBOSS observation time, and kept only quasars in the redshift range $0.8 < z < 2.2$.

\begin{figure}[ht]
    \epsscale{1.18}
    \plotone{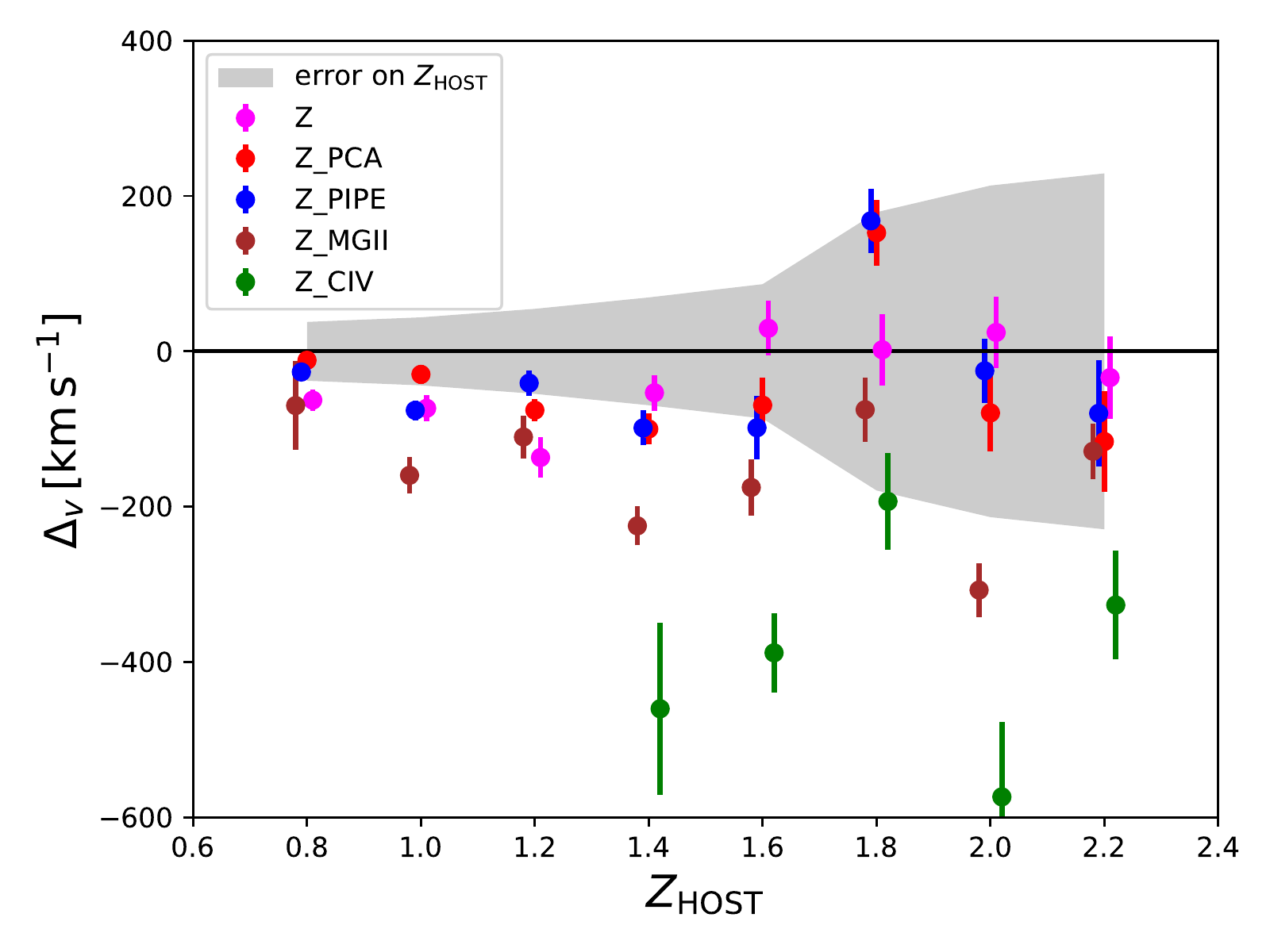}
    \caption{The velocity difference between measured redshift and the host redshift, \zsys, as a function of \zsys\ for different redshift tracers. The grey area represents the 1-$\sigma$ uncertainty for \zsys. The velocity difference is defined in Eqn.~\ref{eqn:vdiff}, though the redshift difference is used in place of the absolute redshift difference to characterize systematic red- or blueshifting of the velocity difference. ``Host'' does not imply this is an unbiased estimate of the redshift, but an estimate derived from the host galaxies of quasars at low redshift.\label{fig:z_v_sys}}
\end{figure}

We define the velocity difference for redshifts as:
\begin{equation}\label{eqn:vdiff}
    \Delta v = c\times\frac{|Z_{\textrm{\footnotesize{c}}}-Z_{\textrm{\footnotesize{T}}} |}{1 + Z_{\textrm{\footnotesize{T}}}},
\end{equation}
where $Z_{\textrm{\footnotesize{c}}}$ is the redshift under comparison, $c$ is the speed of light, and \zt\ is the redshift baseline used for comparison. In Fig.~\ref{fig:z_v_sys}, \zsys\ is used for \zt, which is compared to various redshifts available in the catalog as a function of this host redshift. Only $Z$ (the ``primary'' redshift detailed in \S\ref{ssec:bestredshift}), \zpca, and \zpipe\ show systematic shifts that are less than the uncertainty on the host redshift itself. Our study also confirmed that $Z_{\textrm{\footnotesize{\ion{Mg}{2}}}}$ is the least biased broad emission line redshift estimate as previously reported in, e.g., \citet{shen2016}.

\begin{figure}[ht]
    \epsscale{1.18}
    \plotone{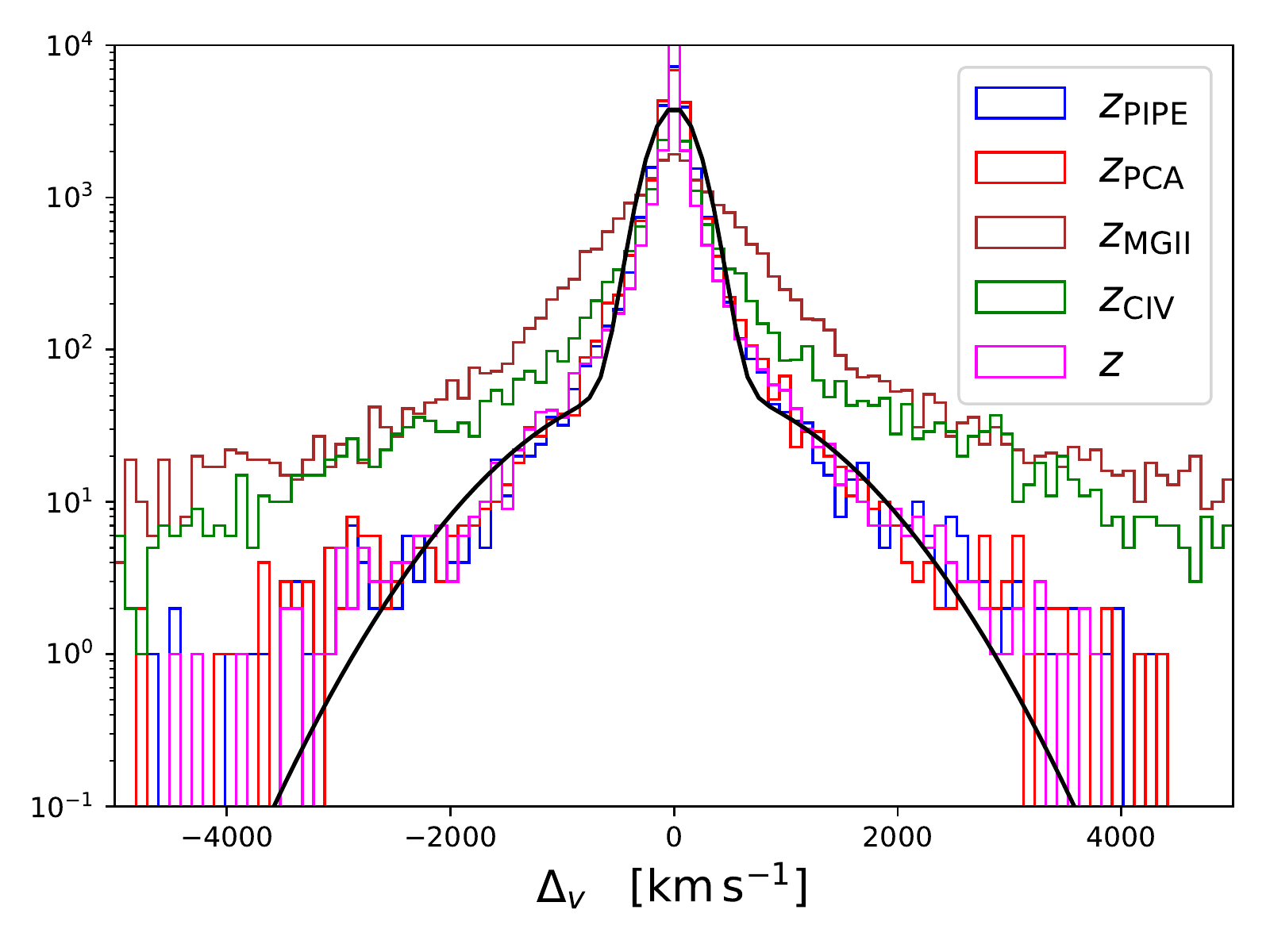}
    \caption{The distribution of the velocity difference for the same redshift tracers seen in Fig.~\ref{fig:z_v_sys}. The distributions are centered on 0\,\kms\ within statistical uncertainty. The double Gaussian distribution (solid black line) was empirically derived. The velocity difference is defined in Eqn.~\ref{eqn:vdiff} using the redshift difference in place of the absolute difference.\label{fig:deltav_sys}}
\end{figure}

In addition to systematic uncertainty, each measured redshift contains a statistical precision that we can estimate using duplicate observations. These duplicate observations are made possible in overlapping plates when unused fibers are assigned to quasars with previous observations. Fig.~\ref{fig:deltav_sys} shows the offset in velocity between two redshift measurements of the same quasar. The statistical precision of \zpca\ and \zpipe\ show a similar behaviour with no significant redshift dependence for the eBOSS redshift range, and show a statistical uncertainty of $\sim 300$\,\kms. A more detailed discussion of statistical uncertainties in \zpca\ and \zpipe\ can be found in Appendix~\ref{asec:zstatuncertainties}.

\begin{deluxetable*}{>{\bf}lCCCCCC}[ht!]
\tablecaption{Velocity Differences by \texttt{SOURCE\_Z} \label{tab:vdiff}}
\tablecolumns{7}
\tablehead{
\colhead{\texttt{SOURCE\_Z}} & \colhead{Sample Size} & \colhead{Acceptable $\Delta v$} & \colhead{Catastrophic Failures} & \colhead{Catastrophic Failure Rate} & \colhead{Median $\Delta v$} & \colhead{MAD $\Delta v$} \\
\colhead{} & \colhead{} & \colhead{$\Delta v \leq 3000$\,\kms} & \colhead{$\Delta v > 3000$\,\kms} & \colhead{} & \colhead{\kms} & \colhead{\kms}
}
\startdata
VI & 319{,}518 & 314{,}392 & 5{,}126 & 1.60\,\% & 216.2 & 165.5\\
PIPE & 355{,}948 & 354{,}713 & 1{,}235 & 0.35\,\% & 48.5 & 33.9 \\
DR7Q\_SCH & 11{,}124 & 11{,}002 & 122 & 1.10\,\% & 148.5 & 115.4\\
DR6Q\_HW & 63{,}133 & 62{,}641 & 492 & 0.78\,\% & 130.5 & 87.8\\
DR12Q & 18 & 18 & 0 & 0.00\,\% & 170.0 & 123.1\\
\hline
TOTAL & 749{,}741 & 742{,}766 & 6{,}975 & 0.93\,\% & 93.0 & 73.5\\
\enddata
\tablecomments{MAD $\Delta v$ is the median absolute deviation of the velocity difference.}
\end{deluxetable*}

For the catalog as a whole, we define quasars in DR16Q as being a ``catastrophic failure'' if they have a velocity difference from a reference redshift of $\Delta v > 3000$\,\kms. As the \citet{shen2015} quasar sample is much smaller than DR16Q, we do not have `host' redshifts to calculate catastrophic failures for the entire catalog. We define a different redshift, \zpca, as \zt\ for comparing the catastrophic failure rate of redshifts from various sources in the \texttt{Z} column. For a detailed discussion of the use of \zpca\ in the clustering catalog, see \citet{ross20a}. The comparisons of the various sources to \zpca\ can be found in Table~\ref{tab:vdiff}, which also includes a comparison to $Z$ for a ``full sample'' of quasars in DR16Q. The full sample consists only of quasars where $0<Z\leq5.0$ and $0<$~\zpca~$\leq5.0$, and each source subsample was drawn from this greater set.

\begin{figure*}[ht]
    \epsscale{1.0}
    \plottwo{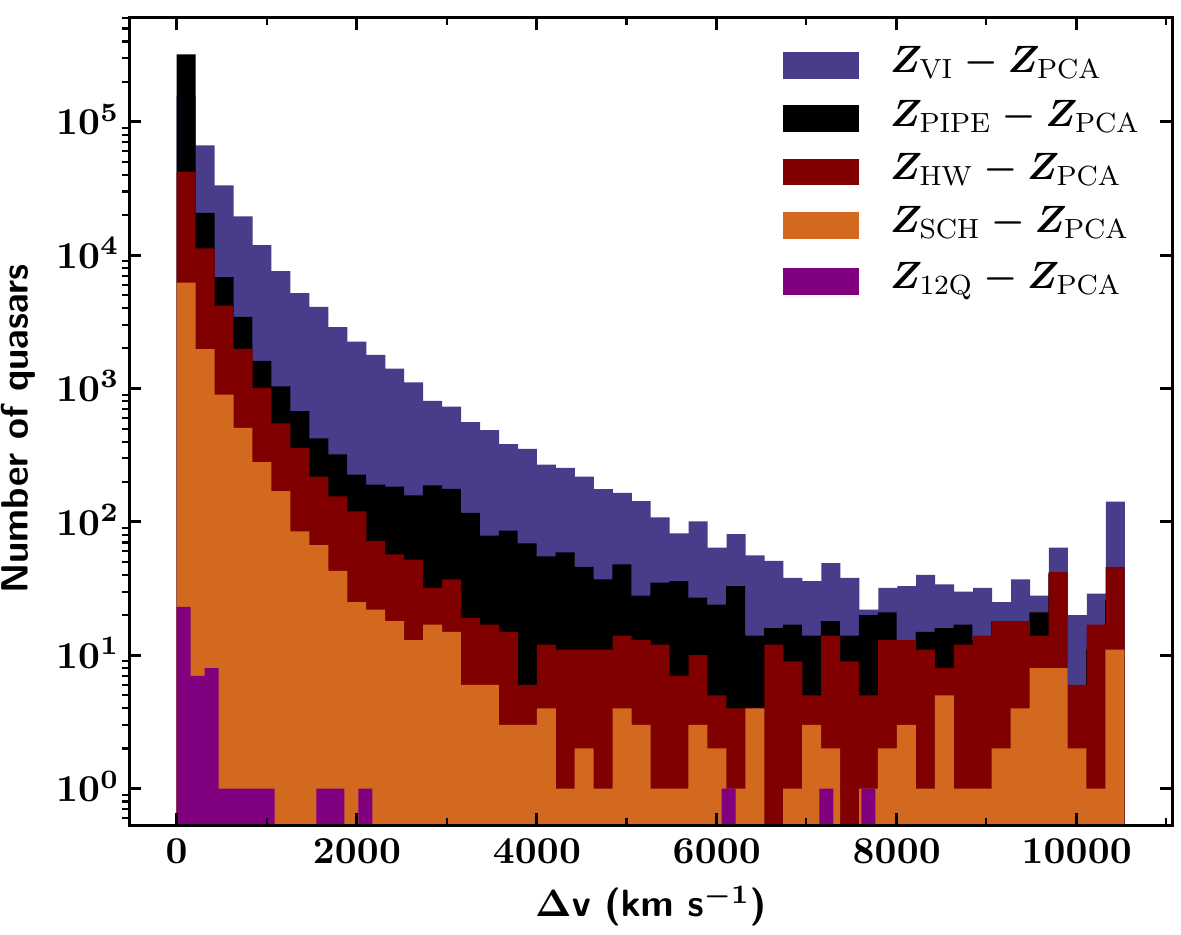}{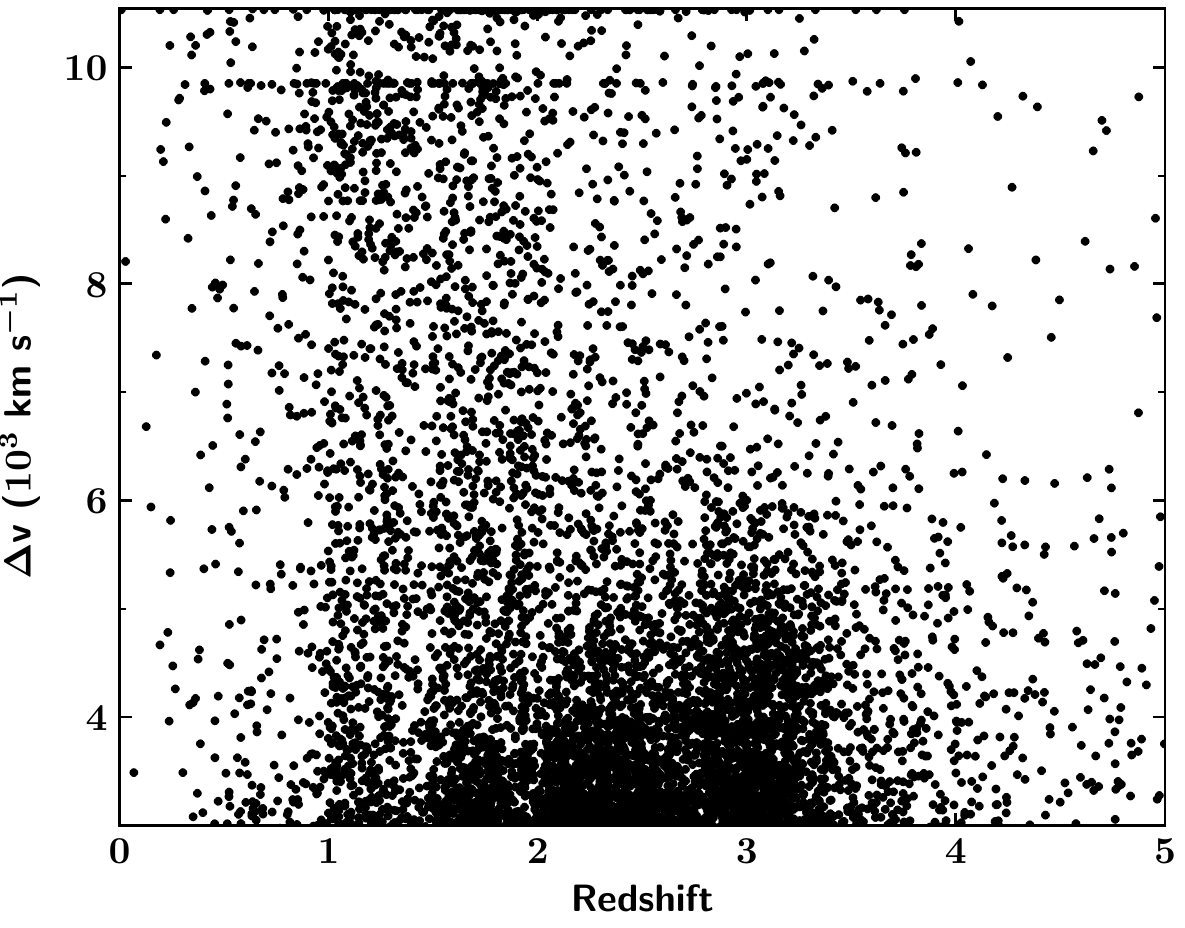}
    \caption{\textit{Left:} The distribution of velocity differences between different redshift values and \zpca~for quasars, where $\Delta v$ is defined by Eqn.~\ref{eqn:vdiff}. The five distributions overlap; they are not vertically stacked. \textit{Right:} The velocity difference as a function of the PCA redshift for catastrophic failures. A catastrophic failure is defined as $\Delta v > 3000$\,\kms. The horizontal lines that appear around 9{,}900\,\kms\ and 10{,}500\,\kms\ are artifacts of edge effects due to the PCA-fitting algorithm binning method. \label{fig:vdiff}}
\end{figure*}

The full distribution of $\Delta v$ for the samples in Table~\ref{tab:vdiff} can be seen in the left-hand panel of Fig.~\ref{fig:vdiff}. The right-hand panel shows $\Delta v$ as a function of \zpca\ for catastrophic failures in the full sample. The lines appearing near 9{,}900\,\kms\ and 10{,}500\,\kms\ are artifacts of edge effects due to how the PCA-fitting algorithm bins redshift ranges.

Over the entire sample (marked ``TOTAL'' in Table~\ref{tab:vdiff}) the catastrophic failure rate for the \texttt{Z} column is 0.93\%. It should be noted that the PCA redshift, itself, is likely not an ideal estimate of a quasar's host redshift. In addition, the PCA-fitting algorithm shares many properties with the SDSS pipeline (\texttt{SOURCE\_Z}=`PIPE'), so \texttt{Z\_PIPE} and \texttt{Z\_PCA} might be expected to be similar.

\section{Damped Lyman Alpha and Broad Absorption Line systems}\label{sec:dlabal}
As was the case for previous SDSS quasar catalogs, DR16Q includes information about DLA systems (henceforth DLAs) and BAL quasars (henceforth BALs). For DR16Q, we used automated processes rather than visual inspections to identify DLAs and BALs, but BALs identified in previous visual inspection campaigns still retain that classification independent of the processes described in this section.

\subsection{Damped Lyman Alpha systems}\label{ssec:dla}
We identified DLAs in DR16Q quasar spectra using the algorithm described in \cite{parks2018}, which is based on a convolutional neural network \citep[CNN; see][for full details]{parks2018}.

We only classified DLAs in spectra with $2 \leq$~\zpca~$\leq 6$, the redshift range over which spectra contained enough pixels to reliably identify DLAs. We thus reduced the input set to 270{,}315 sightlines. If multiple observations were available for one object in the \texttt{spAll-v5\_13\_0} file, we used the stacked spectrum of all good observations as input to the DLA finder. We identified bad spectra using the \texttt{ZWARNING} parameter (see, e.g., \S\ref{ssec:autoclass}). If \texttt{ZWARNING} was {\tt SKY}, {\tt LITTLE\_COVERAGE}, {\tt UNPLUGGED}, {\tt BAD\_TARGET}, or {\tt NODATA}, we did not use the associated observation in the stack. Following \citet{parks2018}, we provide a very pure sample that has a confidence parameter of more than 0.9, and logarithms of the column densities of more than 20.3. Our final sample contains 39{,}514 DLAs in 35{,}686 sightlines with an average logarithmic column density of 20.606. For each quasar spectrum, we provide the list of identified DLAs with the absorption redshifts (\texttt{Z\_DLA}), the logarithms of the column density (\texttt{NHI\_DLA}), and the confidence parameters (\texttt{CONF\_DLA}). The size of each list is the same and corresponds to the maximum number of identified DLAs in any spectrum. If all parameters are set to -1 then no DLA was detected. More information about the efficiency and purity of the aforementioned algorithm is forthcoming in Chabanier et al. (in prep).

\subsection{Broad Absorption Line systems}\label{ssec:bals}
We identified BALs in all quasars at $1.57 \leq z \leq 5.6$ using an algorithm that looks for absorption troughs that would represent either blueshifted \civ\ or \siiv\ features. We performed a $\chi^2$ fit of an unabsorbed quasar model to each spectrum, looked for differences between the model and spectrum that would represent absorption, masked out these regions, and iterated these steps until no new absorption features are identified. This procedure is very similar to the method described by \citet{guo2019} to prepare DR14 quasar spectra for input to their CNN, and used the same five principal components. We performed this fit over the rest-frame wavelength range 1260\,\AA\ to 2400\,\AA\ when possible, and a shorter range for the lowest and highest redshift quasars. If the $\chi^2$ value of this fit was worse than the $\chi^2$ value of the best pipeline fit, we used that pipeline fit instead. We then normalized the spectrum by the best model.

The algorithm measured the commonly used ``BALnicity index'' (BI) proposed by \citet{weymann1991} and the intrinsic absorption index (AI) proposed by \citet{hall2002}. BI was computed using,
\begin{equation}\label{eqn:bi}
    {\rm BI} = - \int^{3000}_{25000}{\left[1-\frac{f(v)}{0.9}\right]C(v){\rm d}v},
\end{equation}
where $C(v) = 0$, unless $1-f(v)/0.9$ is continuously positive over a velocity interval $\Delta v \geq 2000$\,\kms, in which case, $C(v) = 1$. $f(v)$ is the normalized flux density as a function of velocity blueshift from the \civ\ or \siiv\ emission-line center. AI was computed by,
\begin{equation}\label{eqn:ai}
    {\rm AI} = - \int^{0}_{25000}{\left[1-\frac{f(v)}{0.9}\right]}C(v){\rm d}v,
\end{equation}
where $C(v)$ and $f(v)$ have the same definition as Eqn.~\ref{eqn:bi}. DR16Q contains any BI and AI measurements for both \civ\ and \siiv\ for all quasars with $1.57 \leq z \leq 5.6$. These are \texttt{BI\_CIV}, \texttt{AI\_CIV}, \texttt{BI\_SIIV}, and \texttt{AI\_SIIV}. We also provide error estimates for each quantity: \texttt{ERR\_BI\_CIV}, \texttt{ERR\_AI\_CIV}, \texttt{ERR\_BI\_SIIV}, \texttt{ERR\_AI\_SIIV}. All BI, AI, and error estimates are in units of \kms.

We assigned a BAL probability \texttt{BAL\_PROB} to each quasar based on the statistical significance of the troughs associated with the \civ\ line, and the quality of the $\chi^2$ fit to the quasar continuum. The \texttt{BAL\_PROB} values are one of four, discrete values, and are based on visual inspection of large numbers of quasars whose BI values and continuum fits have different levels of statistical significance. For quasars with good continuum fits, we assigned \texttt{BAL\_PROB} = 1 to cases where \texttt{BI\_CIV} is more than ten times the uncertainty in this quantity. Based on our visual inspections, these are all unambiguous cases. Quasars with a less significant BI measurement, but whose \texttt{AI\_CIV} is more than ten times the uncertainty, were assigned \texttt{BAL\_PROB} = 0.95. These are nearly all unambiguous BALs as well. If \texttt{BI\_CIV} is zero, but AI is similarly significant, we assigned \texttt{BAL\_PROB} = 0.9. These are nearly all BALs, but their BI value is zero where the trough is under 2000\,\kms\ wide and/or the trough extends closer to the line center than 3000\,\kms. Other quasars with a less significant BI measurement were assigned \texttt{BAL\_PROB} = 0.75, and quasars with a \texttt{BI\_CIV} value of zero and a less significant AI value were assigned \texttt{BAL\_PROB} = 0.5. We used similar criteria to assign \texttt{BAL\_PROB} to quasars with poorer continuum fits, although with generally lower probability values. For a small number of quasars, the continuum fit failed, and these quasars were not assigned a value for \texttt{BAL\_PROB}. The remaining quasars were assigned \texttt{BAL\_PROB} = 0.

\section{Summary of quasar characteristics}\label{sec:quasarsummary}
DR16Q comprises two catalog files: a superset of objects targeted as quasars by SDSS-I/II/III/IV and a quasar-only set selected from that superset. The superset contains 1{,}440{,}615 observations and includes quasars from BOSS and eBOSS, legacy quasars added from DR7Q, and serendipitous discoveries added from DR12Q. The quasar-only catalog contains 750{,}414 quasars, where 225{,}082 are new quasars observed since DR14Q. Quasars appearing in DR16Q can be identified using the unique combination of plate, MJD, and fiber ID. The plate, MJD, and fiber ID of ``duplicate observations'' have also been back-populated into the superset for user convenience, though each observation still appears in its own record. We have also marked which quasars are considered primary in the \texttt{PRIM\_REC} field, which only appears in the superset. Due to catalog construction constraints in DR7Q, there are possible additional duplicate observations from SDSS-I/II that are not included in DR16Q. DR16Q represents the largest catalog of quasars to date taken from SDSS data and covers 9{,}376\,\degsqr\ of the sky, with an average on-sky surface density of $\sim80$\,\degdeg. DR16Q also presents the largest variety of redshift estimates, to date, for each quasar in an SDSS quasar catalog and is the first such catalog to include the Gaia DR2 data for known quasars.

\begin{figure*}[ht]
    \epsscale{1.0}
    \plottwo{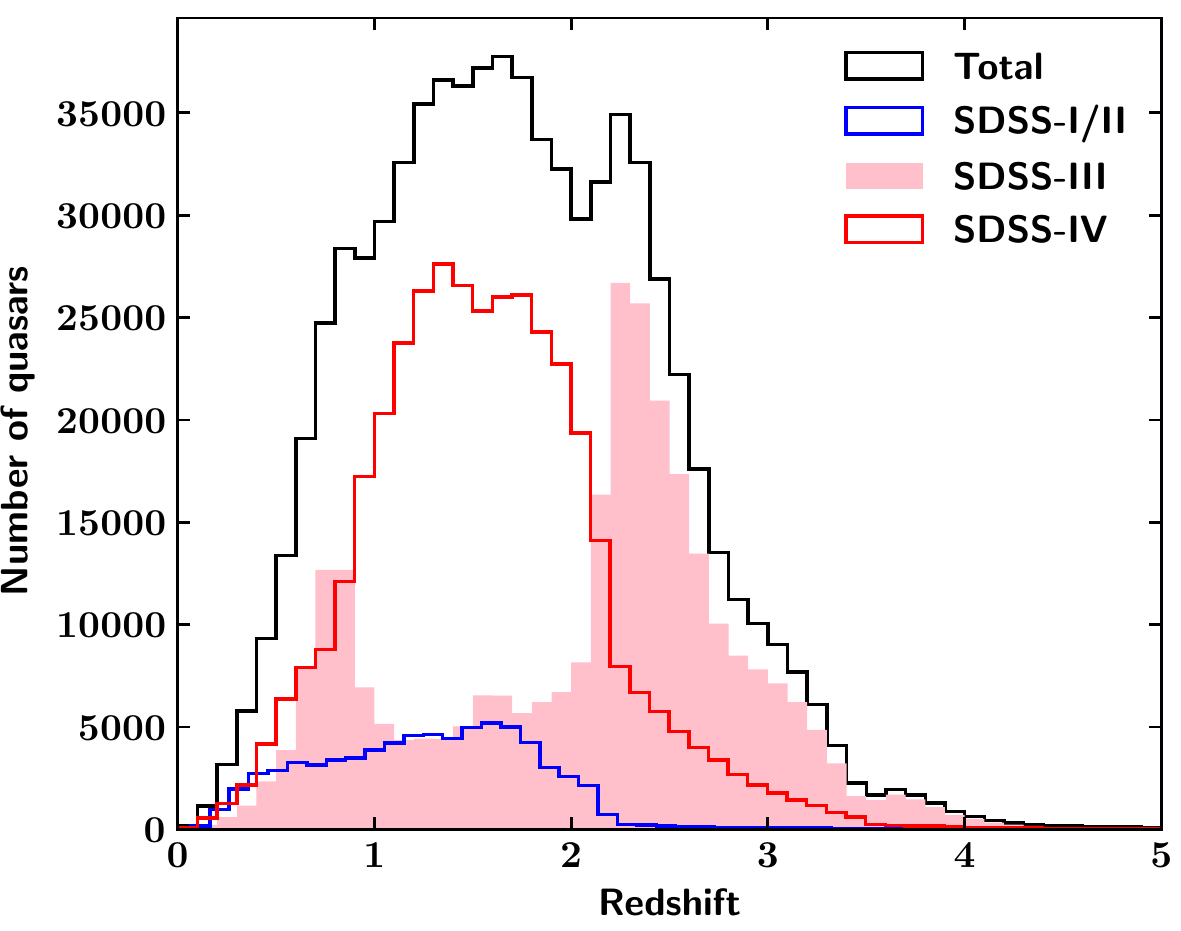}{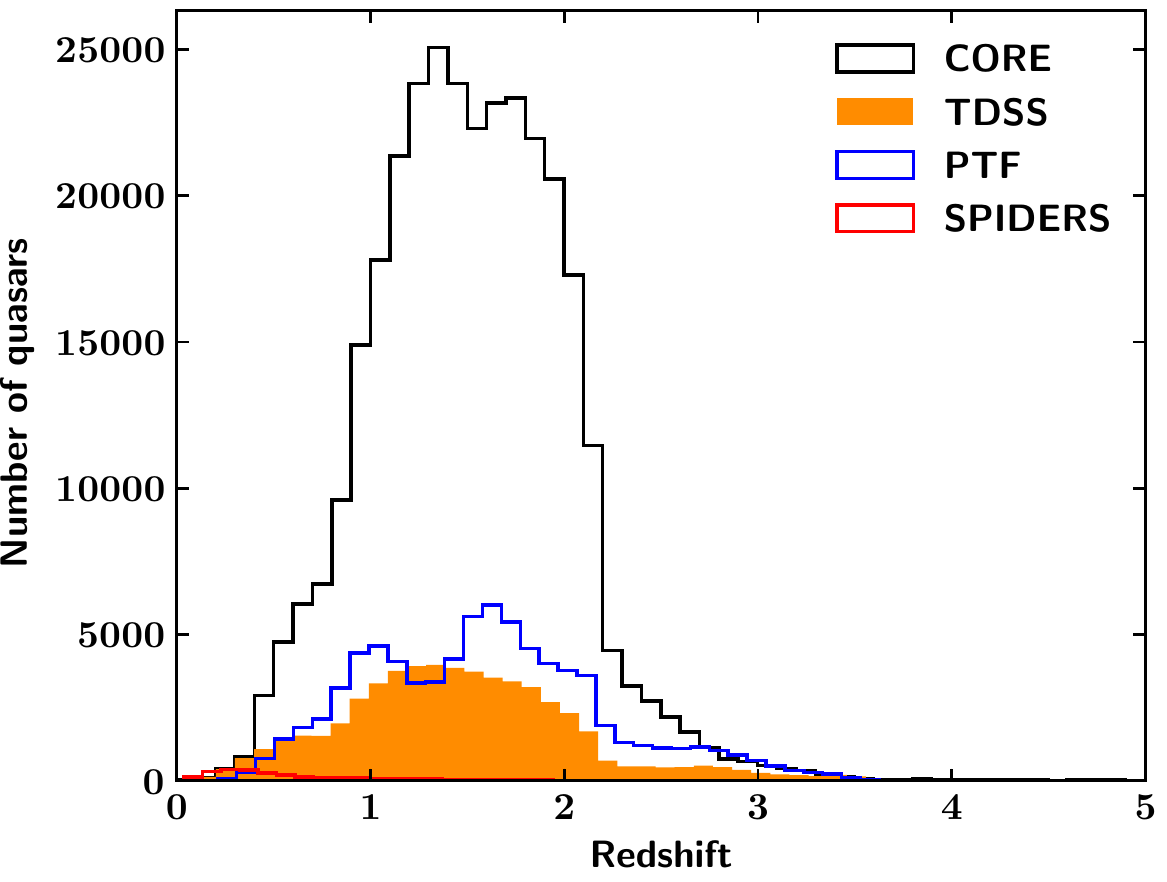}
    \caption{\textit{Left:} The number of DR16Q quasars as a function of the PCA redshift, \zpca, for different SDSS spectroscopic campaigns. \textit{Right:} The number of SDSS-IV/eBOSS quasars in DR16Q, separated by eBOSS subprogram as a function of \zpca. \textit{Both:} In both panels only quasars where $0 <$ \zpca\ $\leq 5$ were used.\label{fig:zhisto}}
\end{figure*}

\begin{figure}[ht]
    \epsscale{1.18}
    \plotone{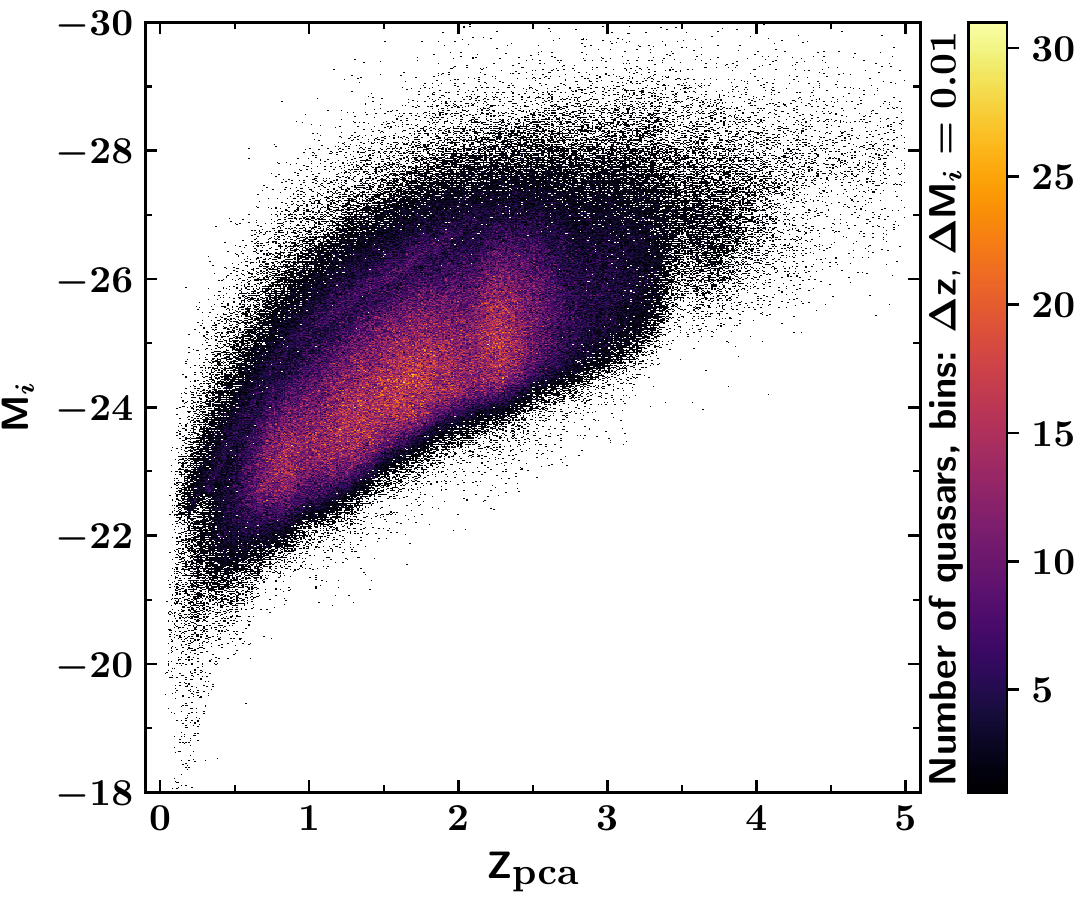}
    \caption{The distribution of absolute $i$-band magnitudes, $M_{i}$[$z=2$], as a function of the PCA redshift, \zpca. Magnitudes and redshifts were separated into bins of size $\Delta$\zpca\ $=0.01$ and $\Delta M_{i}=0.01$. See \S\ref{sec:description} for details on how $M_{i}$[$z=2$] was calculated. Only quasars that were detected at 1-$\sigma$ in the {\it i}-band are included in the plot.
    \label{fig:absmag}}
\end{figure}

The left-hand panel of Fig.~\ref{fig:zhisto} presents the redshift distribution of the number of observed quasars for each SDSS campaign as they appear in the DR16Q quasar-only catalog. The number of observed quasars increased dramatically during the SDSS-III/BOSS campaign, which targeted fainter quasars at higher redshift. The right-hand panel of Fig.~\ref{fig:zhisto} shows the redshift distribution of only the SDSS-IV/eBOSS quasars broken down by eBOSS subprogram. Quasars in each panel use the PCA-derived redshift for consistency, and are limited to quasars where $0 < $ \zpca\ $\leq 5$. It is clear that the most SDSS quasars are contributed by SDSS-IV and that by far the most SDSS-IV quasars are contributed by the eBOSS {\tt CORE} sample. The luminosity space of the quasars in DR16Q is shown in Fig.~\ref{fig:absmag}, as a function of the same PCA redshifts.


\begin{figure*}[ht]
    \epsscale{1.0}
    \plotone{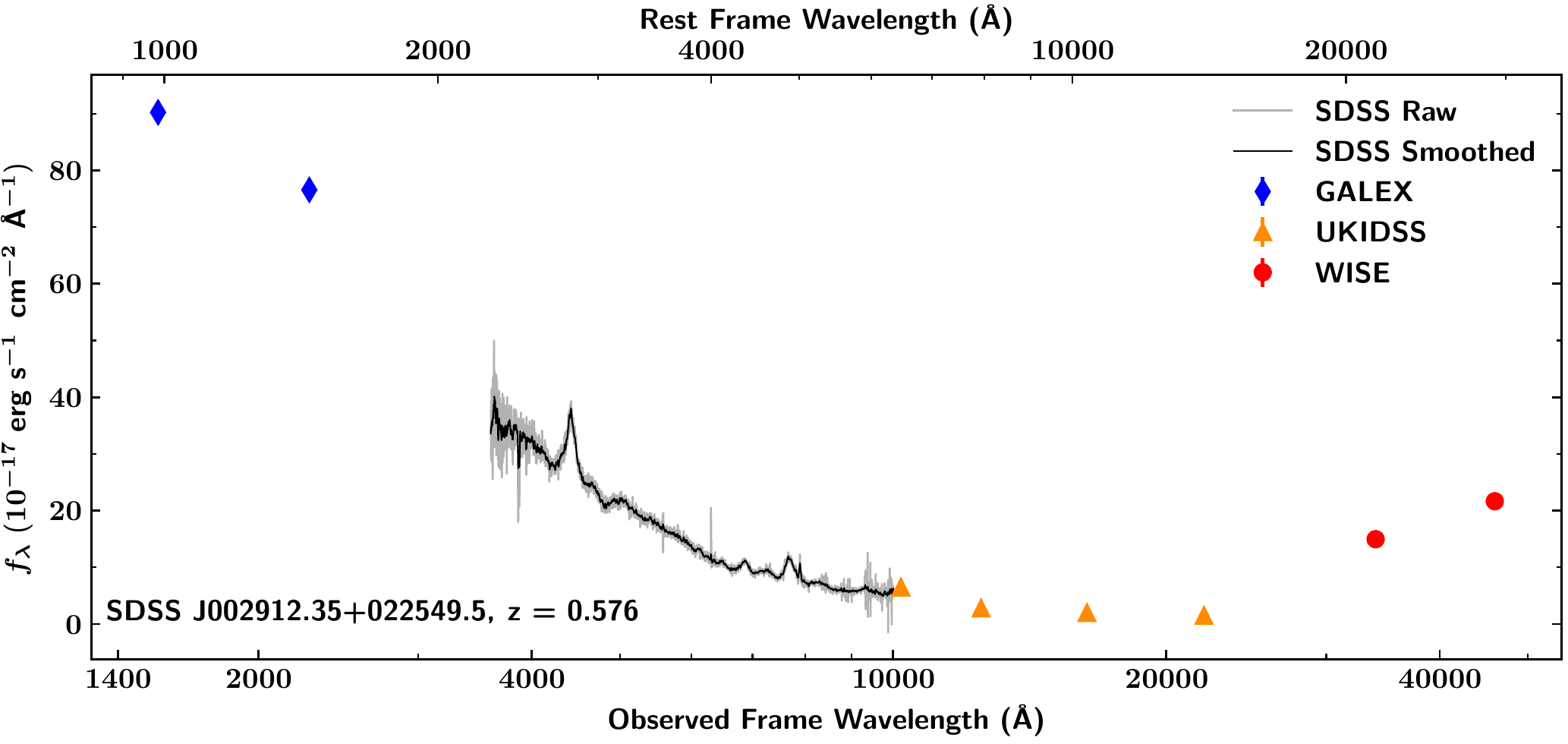}
    \caption{The spectral energy distribution for the quasar SDSS\,J002912.35+022549.5 (plate 7855, MJD 57011, fiber ID 530). Data are included from GALEX, SDSS, UKIDSS, and WISE. The SDSS spectrum (gray) has been smoothed (black) using a boxcar with a width of $8.4$\,\AA\ (10 pixels). Error bars for GALEX, UKIDSS, and WISE data are included, but are too small to be visible at this scale. In the SDSS spectrum the \mgii\ $\lambda2799$, \hb, and \hg\ emission lines can be clearly distinguished. SDSS\,J002912.35+022549.5 also has data in DR16Q from FIRST, ROSAT/2RXS, and Gaia.
    \label{fig:sed}}
\end{figure*}

\section{Multi-wavelength Data}\label{sec:multiwave}
DR16Q includes force-photometered or cross-matched data from: the Galaxy Evolution Explorer (GALEX, \citealt{galex_paper}), the UKIRT Infrared Deep Sky Survey (UKIDSS, \citealt{ukidss_paper}), the Wide-Field Infrared Survey Explorer (WISE, \citealt{wise_mission_paper}), the FIRST radio survey \citep{first_paper}, the Two Micron All Sky Survey (2MASS, \citealt{2mass_paper}), the Second ROSAT All-Sky Survey (2RXS, \citealt{2rxs_paper}), the Third XMM-Newton Serendipitous Source Catalog \citep{xmm_paper}, and Gaia data release 2 \citep{gaiadr2_paper}. In this section, we describe how we incorporated these multi-wavelength data into DR16Q. For GALEX, UKIDSS, and WISE we cross-matched to the data releases that were actually used to {\em target} quasars in BOSS and eBOSS. These multi-wavelength data, then, are particularly useful for trying to recreate BOSS/eBOSS targeting. As FIRST, 2MASS, ROSAT, XMM-Newton, and Gaia were not used in BOSS/eBOSS targeting, we chose to cross-match to the most recent data set available during catalog construction. The data set used for each external survey is explicitly defined in that survey's subsection.

A summary of these external surveys can be found in Table~\ref{tab:externalsummary}, with more detailed information in the appropriate subsections below. In Fig.~\ref{fig:sed} we demonstrate the utility of these data by providing the spectral energy distribution (SED) of a quasar that has extensive multi-wavelength coverage in DR16Q.

\begin{deluxetable*}{llC}[ht!]
\tablecaption{Summary of Multi-wavelength Surveys \label{tab:externalsummary}}
\tablecolumns{3}
\tablewidth{0pt}
\tablehead{
\colhead{Survey} & \colhead{Passband Centers} & \colhead{Number of DR16Q Sources}
}
\startdata
GALEX (\S\ref{ssec:galex}) & UV (NUV, 1350--1750\,\AA; FUV (1750--2750\,\AA) & 646{,}041 \\
UKIDSS (\S\ref{ssec:ukidss}) & Infrared ({\it Y}, $1.02$\,\micm; {\it J}, $1.25$\,\micm; {\it H}, $1.63$\,\micm; {\it K}, $2.20$\,\micm) & 151{,}362 \\
WISE (\S\ref{ssec:wise}) & Infrared (W1, $3.4$\,\micm; W2, $4.6$\,\micm) & 747{,}962 \\
FIRST (\S\ref{ssec:firstradio}) & Radio (20\,\cm) & 21{,}843 \\
2MASS (\S\ref{ssec:2mass}) & Infrared ({\it J}, $1.25$\,\micm; {\it H}, $1.65$\,\micm; $K_s$, $2.16$\,\micm) & 18{,}115 \\
ROSAT/2RXS (\S\ref{ssec:rosat}) & X-ray (0.5--2.0\,\kev) & 11{,}545 \\
XMM-Newton (\S\ref{ssec:xmm}) & X-ray (Soft, 0.2--2.0\,\kev; Hard, 2.0--12.0\,\kev) & 18{,}138 \\
Gaia (\S\ref{ssec:gaia}) & Optical (G, 3300--10{,}500\,\AA; BP, 3300--6800\,\AA; RP, 6300--10{,}500\,\AA) & 469{,}786 \\
\enddata
\end{deluxetable*}

\subsection{GALEX}\label{ssec:galex}
As in past SDSS quasar catalogs, DR16Q includes data from GALEX Data Release 5, force-photometered at the location of SDSS DR8 imaging sources \citep{dr8_paper}. We present data from both GALEX bands: NUV (1350--1750\,\AA) and FUV (1750--2750\,\AA). A total of 646{,}041 objects have a non-zero flux in either the NUV or FUV band, 552{,}025 (431{,}431) have a positive NUV (FUV) flux, and 386{,}642 objects have a positive flux in both bands. All fluxes are reported in nanomaggies\footnote{See \url{https://www.sdss.org/dr16/algorithms/magnitudes/\#Fluxunits:maggiesandnanomaggies}}, where 1 nanomaggy $\approx 3.631\times10^{-6}$\,Jy.

\subsection{UKIDSS}\label{ssec:ukidss}
Similarly to GALEX, DR16Q includes SDSS-DR8-based imaging force-photometered in the four UKIDSS bands (with central wavelengths): {\it Y} ($1.02$\,\micm), {\it J} ($1.25$\,\micm), {\it H} ($1.63$\,\micm), and {\it K} ($2.20$\,\micm) bands. UKIDSS data was taken before March 2011 and was released as UKIDSS DR1--DR9 \citep[see \S4.5.5 of][]{dr10q_paper}. We provide the fluxes and errors for each of these bands in units of \wmh.

DR16Q contains 151{,}362 measurements from the UKIDSS area (which is smaller than the SDSS footprint). [150{,}147, 149{,}629, 149{,}502 and 150{,}288] sources have a positive flux in [{\it Y}, {\it J}, {\it H}, and {\it K}] respectively. 146{,}500 sources have a positive flux in all four bands.

\subsection{WISE}\label{ssec:wise}
eBOSS targeting was, in part, based on data taken from the WISE W1 ($3.4$\,\micm) and W2 ($4.6$\,\micm) bands \citep{eBOSS_target_select}. Many of the eBOSS quasar targets had fluxes below the WISE detection limit used for the AllWISE Data Release, so forced photometry was applied to custom ``unWISE'' stacks \citep{unwise_coadds}, as described in \citet{lang2014}.\footnote{For more information see: \url{https://www.sdss.org/dr15/data_access/value-added-catalogs/?vac_id=wise-forced-photometry}}

DR16Q contains 747{,}962 objects with a positive flux in either the W1 or W2 bands. There are 744{,}835 (741{,}227) objects with a positive flux in the W1 (W2) band, and 739{,}093 objects with a positive flux in both bands. Quasars in DR16Q that were not identified using SDSS imaging do not have ``unWISE'' forced photometry. There are 1001 such objects in DR16Q and each has a flux set to -1 in both bands.

\subsection{FIRST}\label{ssec:firstradio}
Sources in DR16Q are matched to the December 2014, version of the FIRST\footnote{\url{http://sundog.stsci.edu/first/catalogs.html}} catalog \citep[e.g.][]{first_2015}, using a $2.0\arcsec$ radius. Our reported \texttt{FIRST\_FLUX} corresponds to the peak flux density, in \mjy, at $20$\,\cm\ wavelength. These values include the added $0.25$\,\mjy\ ``CLEAN'' bias described in \citet{first_data_paper}. We also report the FIRST signal-to-noise ratio of the peak flux using,
\begin{equation}\label{eqn:first_sn}
    \textrm{S/N} = \frac{f_{\textrm{peak}} - 0.25}{f_{\textrm{rms}}}.
\end{equation}
The catalog contains 21{,}843 matches to FIRST radio sources.

\subsection{2MASS}\label{ssec:2mass}
DR16Q includes data from the 2MASS All-Sky Point Source Catalog, released March 2003, in three bands (with central wavelengths): {\it J} ($1.25$\,\micm), {\it H} ($1.65$\,\micm), and $K_s$ ($2.16$\,\micm). Catalog objects were matched to 2MASS sources within $2.0\arcsec$, and we include the Vega magnitude, magnitude uncertainty, signal-to-noise ratio, and photometric read flag for each object. There are a total of 18{,}115 2MASS matches in DR16Q. The magnitude limits of the 2MASS survey ($J < 15.8$, $H < 15.1$, and $K_s < 14.3$) explain the lower detection rate compared to WISE and UKIDSS.

\subsection{ROSAT/2RXS}\label{ssec:rosat}
Unlike the other surveys mentioned in this section, a single coordinate-match with the ROSAT/2RXS \citep{2rxs_paper} and XMMSL2\footnote{For SL2 specifics see \url{https://www.cosmos.esa.int/web/xmm-newton/xmmsl2-ug}} \citep{saxton2008} would not provide a reliable association. This is because both surveys have a large positional uncertainty, as can be seen in Fig.~1 of \citet{salvato2018}. In particular, 95\% of the 2RXS sources have a 1-$\sigma$ positional uncertainty smaller than 29\arcsec. As detailed in \citet{dwelly2017}, \citet{salvato2018} and \citet{comparat2019}, a better approach is to construct a match between X-ray sources and AllWISE sources via a Bayesian approach (using a code called {\tt Nway}). Here, we simply matched the coordinates of DR16Q within 1\arcsec\ to the AllWISE counterparts of the X-ray sources presented in \citet{salvato2018}.

In DR16Q there are a total of 11{,}545 matched objects, and we include the 2RXS right ascension and declination, and the source flux and flux error in the 0.5--2.0\,\kev\ band.

\subsection{XMM-Newton}\label{ssec:xmm}
DR16Q includes cross-matches to the two most complete X-ray catalogs to date: XMMPZCAT\footnote{See \url{http://xraygroup.astro.noa.gr/Webpage-prodex/xmmfitcat_access.html}} \citep{ruiz2018} and 3XMM-DR8\footnote{See \url{http://xmmssc.irap.omp.eu/Catalogue/3XMM-DR8/3XMM_DR8.html}}, which are both based on the third version of the XMM-Newton Serendipitous Source Catalog \citep{xmm_paper}.

XMMPZCAT \citep{ruiz2018} is a catalog of photometric redshifts for X-ray sources, which was created using a machine learning algorithm \citep[\texttt{MLZ-TPZ}; detailed in][]{carrasco2013}. XMMPZCAT provides X-ray positions and redshifts for about 100{,}000 sources. We cross-matched DR16Q and XMMPZCAT, with a matching radius of 10\arcsec. Despite the large matching radius, 59\% of the sources were within 1\arcsec. DR16Q includes more than 15{,}800 sources with X-ray counterparts in XMMPZCAT, of which $\sim9\%$ are estimated to be contaminants due to chance superpositions.

XMMPZCAT is based on 3XMM-DR6 (the sixth release of the 3XMM catalog). So, in addition, we matched DR16Q to the more recent 3XMM-DR8 catalog using a $5.0\arcsec$ radius. We restricted this match to areas not covered by 3XMM-DR6 or to sources without XMMPZCAT counterparts. DR16Q contains an additional $\sim 2300$ X-ray counterparts from 3XMM-DR8.

For each of the 18{,}138 DR16Q quasars with XMM-Newton counterparts, we report the soft (0.2--2.0\,\kev), hard (4.5--12.0\,\kev), and total (0.2--12.0\,\kev) fluxes with associated errors. These were computed as the weighted average of all detections in the three XMM-Newton cameras (MOS1, MOS2, PN). The X-ray luminosities for the total fluxes were also computed and are provided. These are not absorption-corrected. All fluxes and errors are expressed in \ecs\ and the total luminosity was computed using \texttt{Z}.

\subsection{Gaia}\label{ssec:gaia}
DR16Q includes data from the second data release of Gaia\footnote{\url{https://www.cosmos.esa.int/web/gaia/dr2}}. We matched objects using a $0.5\arcsec$ radius, and present the Vega magnitudes and mean signal-to-noise ratio for the three Gaia bands: G, BP, and RP, as well as the Gaia RA, DEC, parallax, and proper motion (for both RA and DEC). We also include the unique Gaia designation, which is guaranteed to remain unique in future Gaia data releases.

A total of 469{,}786 objects in DR16Q have a match to the Gaia catalog, to limiting magnitudes of $G < 22.0$, $BP < 23.6$, and $RP < 23.2$.

\section{Description of the DR16Q catalog}\label{sec:description}

The DR16Q quasar-only catalog contains 183 columns of information for each quasar in a binary FITS \citep{FITS} table file. The DR16Q superset includes the columns listed below up to column {\bf 98} inclusive, omitting {\tt Z\_LYAWG} and {\tt M\_I} (columns {\bf 53} and {\bf 97} respectively), but adding {\tt PRIM\_REC} (column {\bf 16} in the superset) which has a value of 1 if the record is considered the primary observation for a quasar. The columns in the quasar-only catalog are summarized in Table~\ref{tab:fields}.
\\
\\
Complete DR16Q quasar-only column descriptions follow:\\
\textbf{1.} The SDSS name generated from the RA and DEC for the primary record in the format hhmmss.ss$\pm$ddmmss.s in the J2000 equinox. The 'SDSS\,J' is omitted. Coordinate values are truncated, not rounded.\\
\textbf{2--3.} The right ascension and declination in decimal degrees for the J2000 equinox. \\
\textbf{4--6.} The unique spectroscopic plate number, modified Julian date of spectroscopic observation, and fiber ID. The combination of these three values gives a unique identifier for every spectroscopic observation in SDSS-I/II/III/IV. Where an object was observed multiple times we chose the observation with the most confident visual inspection. If data from visual inspection was not available the observation with the highest \texttt{SN\_MEDIAN\_ALL} was chosen as the primary.\\
\textbf{7.} The automated classification as outlined in \S\ref{ssec:autoclass}. This column can take the values \texttt{GALAXY, QSO, STAR, UNK,} and \texttt{VI}. A classification of \texttt{UNK} corresponds to objects that were added to the superset catalog after the initial automated classification was completed. All of these records were taken from a catalog of visually inspected objects that did not have a targeting bit that appears in Table~\ref{tab:tarbits}.\\
\textbf{8.} The automated classification as detailed in the first part of \S\ref{ssec:autoclass}. This column can take the same set of values as \texttt{AUTOCLASS\_PQN}, and \texttt{UNK} is defined as above. Classifications of \texttt{STAR, GALAXY, QSO} were retained in \texttt{AUTOCLASS\_PQN}. Only objects with a classification of \texttt{VI} in this column could be reclassified in \texttt{AUTOCLASS\_PQN}.\\
\textbf{9.} A numeric flag indicating if an object was identified as a quasar by \qnet. A value of -1 indicates the object was not processed by \qnet, 1 (0) indicates the object was identified as a quasar (non-quasar).\\
\textbf{10.} The redshift of the quasar as computed by \qnet. As outlined in \S\ref{ssec:autoclass}, this was used as a binary discriminant with a threshold set to $\texttt{Z\_QN} = 2.0$. Observations with \texttt{AUTOCLASS\_DR14Q} set to \texttt{VI} that had $\texttt{Z\_QN} < 2.0$ were reclassified in \texttt{AUTOCLASS\_PQN} as \texttt{QSO}.\\
\textbf{11.} A binary flag indicating if the object was selected randomly for visual inspection as detailed in \S\ref{ssec:eboss_vi}. A value of 0 indicates the object was not selected. The randomized subsample was selected from the superset and contained some duplicate observations of quasars. All subsampled 10{,}000 observations were unique. Objects were selected from a larger set where \texttt{EBOSS\_TARGET1} contained the \texttt{QSO1\_EBOSS\_CORE} targeting bit set, or had an observation date (MJD $\geq 56839$) and pipeline redshift $\geq 1.8$\\
\textbf{12.} The redshift from the visual inspection of the randomly selected subsample. Note that this field is included as a possible ``visual inspection'' source for \texttt{Z}.\\
\textbf{13.} The confidence rating for a visually identified redshift. A value of -1 indicates the object was not selected for the subsample. A value of 1 is the lowest confidence and 3 is the highest confidence. A value of 1 typically indicates a spectrum with only one broad emission peak above the noise spectrum.\\
\textbf{14.} A numeric flag indicating whether the automated pipeline correctly identified the classification and redshift with $\Delta v \leq 3000$\,\kms. A value of -1 indicates the record was not included in the random subsample, a value of 0 indicates the pipeline incorrectly identified the classification or redshift, and a value of 1 indicates that both classification and redshift were correct.\\
\textbf{15.} A numeric flag that classifies the record as a quasar for the randomly selected subsample. As in \textbf{13-14}, a -1 value indicates the record wasn't selected.\\
\textbf{16.} A 64-bit integer that identifies BOSS and eBOSS objects in the SDSS photometric and spectroscopic catalogs. Some SDSS-I/II objects have a \texttt{THING\_ID} value, but not all. A value of -1 indicates the record wasn't assigned a \texttt{THING\_ID}.\\
\textbf{17.} The visually identified redshift of the record. A value of -999 indicates a possible blazar and no meaningful redshift from an emission feature can be identified. A value of -1 indicates the record was not visually inspected. See \S\ref{ssec:viredshifts}.\\
\textbf{18.} The confidence rating for the visually identified redshift. Objects with -1 were not visually inspected. Objects with 0 were visually inspected but a confident redshift and classification could not be identified. For other objects a value of 3 is the highest confidence.\\
\textbf{19.} The visually identified classification for the object. Definitions for the numeric values appear in Table~\ref{tab:classperson}. Objects appearing in DR16Q that have values of 1 (Star) or 4 (Galaxy) have low \texttt{Z\_CONF} values. In these cases the automated pipeline classification was used to determine whether the object was a quasar.\\
\textbf{20--24.} The redshift and flag indicating an object is a quasar for objects appearing in DR12Q \citep{dr12q_paper}, DR7Q \citep{dr7q_paper}, and \citet{dr6q_hw_paper}. These values are kept separate from \texttt{Z\_VI} as an object observed more than once may have different redshifts for different pipeline reductions. For DR12Q objects, only the visual inspection redshift was recorded.\\
\textbf{25.} The Hewett and Wild (2010) redshifts applied to DR7Q and taken from the ancillary column of \citet{shen2011}. See column 143 at \url{http://das.sdss.org/va/qso_properties_dr7/dr7.htm} for more information.\\
\textbf{26.} The overall flag indicating if an object is a quasar. See \S\ref{ssec:classresults}.\\
\textbf{27.} The best available redshift taken from \texttt{Z\_VI}, \texttt{Z\_10K}, \texttt{Z\_DR12Q}, \texttt{Z\_DR7Q\_SCH}, \texttt{Z\_DR6Q\_HW}, or \texttt{Z\_PIPE}. Visually identified redshifts with $\texttt{Z\_CONF}>1$ were preferred. \texttt{Z\_DR6Q\_HW} was preferred over \texttt{Z\_DR7Q\_SCH}. In the absence of a confident visual inspection redshift or previous catalog redshift, this value is taken from \texttt{Z\_PIPE}.\\
\textbf{28.} The source for the redshift recorded in column {\bf 27}. DR12Q redshifts were taken from the visual inspection redshifts in that catalog.\\
\textbf{29.} The automated redshift taken from version {\tt v5\_13\_0} of the SDSS reduction pipeline. Objects added from DR7Q that had no BOSS or eBOSS reobservation do not have a pipeline redshift.\\
\textbf{30.} A bit flag for the quality of the pipeline redshift\footnote{\url{www.sdss.org/dr16/algorithms/bitmasks/\#ZWARNING}}.\\
\textbf{31.} A string for the SDSS photometric identification generated from the sky version, run, rerun, camera column, field number, and ID. A blank value indicates the object does not have SDSS photometry and was selected as a target from another source catalog.\\
\textbf{32--34.} The redshift, warning flag, and $\Delta \chi^{2}$ for the  PCA redshift generated by \rvb\ (see \S\ref{ssec:systemvsemission}).\\
\textbf{35--52.} Similar to columns {\bf 32--34}, the \rvb\ pipeline fits for six emission lines: \ha, \hb, \mgii, \ciii, \civ, and \lya.\\
\textbf{53.} A PCA-derived redshift using \rvb\ with spectrum-masking of the \lya\ emission line and forest.\\
\textbf{54--56.} The absorber redshift, absorber column density, and confidence rating for DLA absorbers. No object had more than 5 absorbers. Objects with a value of -1 in the first element were either outside the redshift range for DLA systems, or had no identifiable absorbers.\\
\textbf{57.} The probability an object is a BAL quasar. A -1 indicates the object's redshift was too low to have a detection within the spectroscopic wavelength range.\\
\textbf{58--65.} The BALnicity index (BI), BI uncertainty, absorption index (AI), and AI uncertainty for the \civ\ region and for the \siiv\ region. All columns are in units of \kms.\\
\textbf{66--71.} Targeting bit flags for BOSS, eBOSS, and ancillary BOSS programs (see Table~\ref{tab:tarbits}).\\
\textbf{72--73.} The number of additional spectra for an object from SDSS-I/II (column {\bf 72}) or BOSS/eBOSS (column {\bf 73}). Objects with only one observation will have both values set to 0. SDSS-I/II (SDSS-III/IV) objects have an MJD before (after) 54663.\\
\textbf{74.} The total number of additional spectroscopic observations for an object. This is the sum of columns {\bf 72} and {\bf 73}. Objects with only one observation will have this value set to 0.\\
\textbf{75--78.} The spectroscopic plate, modified Julian date, fiber ID, and spectroscopic instrument for each duplicate observation of an object. As with columns {\bf 4--6}, the combination of plate, MJD, and fiber ID provide a unique reference to each observation. For column {\bf 78}, observations from SDSS-I/II have a value of 1, while observations from BOSS/eBOSS have a value of 2. Objects with multiple duplicate observations are in no particular order.\\
\textbf{79--84.} SDSS photometric information. Combining these columns gives the \texttt{OBJID} in column {\bf 31}.\\
\textbf{85.} The wavelength for which the fiber placement was optimized to account for atmospheric differential refraction \citep{boss_mission}, in \AA.\\
\textbf{86.} The focal plane offset, in \micm, to account for wavelength dependence of the focal plane due to distance from the center of the plate \citep{boss_mission}.\\
\textbf{87--88.} The $x$ and $y$ positions in the focal plane for the spectroscopic fiber hole, in \mm.\\
\textbf{89--90.} The name of the tiling chunk and the tile number for the spectroscopic plate placement on the sky.\\
\textbf{91.} The overall $\textrm{(S/N)}^{2}$ for the spectroscopic plate, taken from the minimum of the two red and two blue cameras for the plate.\\
\textbf{92--93.} The PSF flux and inverse variance for each of the five SDSS bands: \textit{u}, \textit{g}, \textit{r}, \textit{i}, and \textit{z}, in nanomaggies. The fluxes are not corrected for Galactic extinction.\\
\textbf{94--95.} As for columns {\bf 92--93}, but for inverse hyperbolic sine (asinh) AB magnitudes \citep{lupton1999}. Magnitudes are not corrected for Galactic extinction.\\
\textbf{96.} The Galactic extinction values from \citet{galactic_extinction} for the five SDSS bands.\\
\textbf{97.} Absolute \textit{i}-band magnitude corrected for extinction. Calculated from the magnitude and Galactic extinction in columns {\bf 94} and {\bf 96}, using the PCA redshift in column {\bf 32}. K-corrections were normalized to a redshift of 2, using Table~4 of \citet{richards2006}. The following cosmological parameters were used: $H_0 = 67.6$\,\hubu, $\Omega_{\rm M} = 0.31$, and $\Omega_{\Lambda} = 0.69$.\\
\textbf{98.} The median S/N for all good pixels in the five SDSS bands.\\
\textbf{99.} A matching flag for objects in the forced-photometry SDSS-DR8/GALEX catalog. A value of 1 (0) indicates the object was (was not) found in the GALEX catalog.\\
\textbf{100--103.} GALEX FUV and NUV fluxes and inverse variances, expressed in nanomaggies.\\
\textbf{104.} A matching flag for objects in the forced-photometry SDSS-DR8/UKIDSS catalog. A value of 1 (0) indicates the object was (was not) found in the UKIDSS catalog.\\
\textbf{105--112.} The flux density and error in the four UKIDSS bands ({\it Y}, {\it J}, {\it H}, and {\it K}) in units of \wmh.\\
\textbf{113--114.} The W1-band ($3.4$\,\micm) WISE flux and inverse flux variance in nanomaggies and nanomaggies$^{-2}$ respectively. A source with a measured magnitude of 22.5 in the {\em Vega} system would have a flux of 1 nanomaggy. See \citet{lang2014}.\\
\textbf{115--116.} The W1-band magnitude and magnitude uncertainty in the Vega magnitude system. Calculated using columns {\bf 113--114}.\\
\textbf{117.} The W1-band profile-weighted $\chi^{2}$ goodness of fit weighted by the point-spread function in the WISE images (for point sources).\\
\textbf{118.} The W1-band flux signal-to-noise ratio. Calculated using columns {\bf 113--114}.\\
\textbf{119.} The W1-band profile-weighted number of WISE exposures used in the ``unWISE'' coadd images for this source. This is analogous to a depth measurement.\\
\textbf{120--121.} The W1-band profile-weighted flux from other sources within the PSF of this source, and the profile-weighted fraction of flux from external sources for this source.\\
\textbf{122.} The W1-band number of pixels included in the PSF fit for this source.\\
\textbf{123--132.} The W2-band ($4.6$\,\micm) data conforming to the descriptions in columns {\bf 113--122}.\\
\textbf{133.} A matching flag for objects in the FIRST catalog. The maximum matching radius was $2.0\arcsec$. A value of 1 (0) indicates the object did (did not) have a match in the FIRST catalog.\\
\textbf{134--135.} The peak flux density and flux density signal-to-noise ratio at $20$\,\cm\ in units of \mjy. See \S\ref{ssec:firstradio}.\\
\textbf{136.} The matching separation in arcsec between the SDSS and FIRST objects.\\
\textbf{137--138.} The {\it J}-band magnitude and magnitude error from the 2MASS point source catalogs \citep{cutri2003}. Objects used a maximum matching radius of $2.0\arcsec$. Magnitudes are in the Vega system.\\
\textbf{139--140.} The {\it J}-band signal-to-noise ratio and photometric read flag that records the default magnitude origins.\footnote{See \url{https://old.ipac.caltech.edu/2mass/releases/allsky/doc/sec2\_2.html}}\\
\textbf{141--144.} {\it H}-band data conforming to the descriptions in columns {\bf 137--140}.\\
\textbf{145--148.} $K_s$-band data conforming to the descriptions in columns {\bf 137--140}.\\
\textbf{149.} The matching separation in arcsec between SDSS and 2MASS objects.\\
\textbf{150.} The 2RXS ID number designating unique objects.\\
\textbf{151--152.} The right ascension and declination of the 2RXS source in decimal degrees (J2000).\\
\textbf{153--154.} The source flux and source flux error in the 0.5--2.0 \kev\ band in \ecs, as described in \citet{2rxs_paper}. $\textrm{G}=2.4$ and dereddened.\\
\textbf{155.} The matching separation in arcsec between SDSS and 2RXS objects.\\
\textbf{156.} The XMM-Newton Source designation ID.\\
\textbf{157--158.} The right ascension and declination of the XMM-Newton source in decimal degrees (J2000).\\
\textbf{159--160.} The X-ray flux and flux error for the 0.2--2.0\,\kev\ energy band in \ecs.\\
\textbf{161--162.} The X-ray flux and flux error for the 2.0--12.0\,\kev\ energy band in \ecs.\\
\textbf{163-164.} The total X-ray flux and flux error for the full energy range (0.2--12.0\,\kev) in \ecs.\\
\textbf{165.} The total X-ray luminosity for the full energy range (0.2--12.0\,\kev) in \esec.\\
\textbf{166.} The matching separation in arcsec between SDSS and XMM-Newton objects.\\
\textbf{167.} A matching flag for objects that appeared in the Gaia DR2 catalog. Objects were matched using a maximum radius of $0.5\arcsec$. A value of 1 (0) indicates the object did (did not) have a match in the Gaia catalog.\\
\textbf{168.} The unique Gaia designation.\\
\textbf{169--170.} The Gaia DR2 barycentric right ascension and declination in decimal degrees (J2015.5).\\
\textbf{171--172.} The Gaia absolute stellar parallax and parallax error (J2015.5).\\
\textbf{173--174.} The proper motion and standard error of proper motion for right ascension in \masy\ (J2015.5).\\
\textbf{175--176.} The proper motion and standard error for declination in the same form as columns {\bf 173--174}.\\
\textbf{177--178.} The mean magnitude and mean flux over standard deviation for the Gaia G-band in the Vega magnitude system.\\
\textbf{179--180.} BP-band data conforming to the descriptions in columns {\bf 177--178}.\\
\textbf{181--182.} RP-band data conforming to the descriptions in columns {\bf 177--178}.\\
\textbf{183.} The matching separation in arcsec between SDSS and Gaia DR2.

\section{Conclusion}\label{sec:conclusion}
In this paper, we detail DR16Q, the final SDSS-IV/eBOSS quasar catalog. The catalog consists of two subcatalogs: the DR16Q superset containing 1{,}440{,}615 observations targeted as quasars, and the quasar-only set containing 750{,}414 quasars. The quasar-only catalog includes 225{,}082 new quasars observed after the release of DR14Q, a 42\% increase in the catalog size. We estimate the quasar-only catalog to be 99.8\% complete with 0.3--1.3\% contamination. We include automated pipeline redshifts for all quasars observed as part of SDSS-III/IV and confident visual inspection redshifts for 320{,}161 quasars. DLAs and BALs were identified and measured by automated algorithms and the quasar-only catalog includes 35{,}686 DLAs and 99{,}856 spectra with $\texttt{BAL\_PROB} \geq 0.75$. DR16Q includes homogeneous redshifts derived using principal component analysis, including six separate emission line PCA redshifts. Finally, DR16Q includes multi-wavelength matching to GALEX, UKIDSS, WISE, FIRST, 2MASS, ROSAT/2RXS, XMM-Newton and Gaia.

SDSS-IV/eBOSS observations are now complete and SDSS-V \citep{SDSSV} will target far fewer quasars than SDSS-IV. DR16Q is therefore likely to remain the most significant compilation of SDSS quasars for quite some time to come.

\section*{Acknowledgements}\label{sec:ack}

BWL, JNM, and ADM were supported by the U.S.\ Department of Energy, Office of Science, Office of High Energy Physics, under Award Number DE-SC0019022. ANH is supported by The University of Wyoming Science Initiative Wyoming Research Scholars Program and was funded by Wyoming NASA Space Grant Consortium, NASA Grant \#NNX15AI08H.

Funding for the Sloan Digital Sky Survey IV has been provided by the Alfred P.\ Sloan Foundation, the U.S.\ Department of Energy Office of Science, and the Participating Institutions. SDSS-IV acknowledges
support and resources from the Center for High-Performance Computing at the University of Utah. The SDSS web site is www.sdss.org. In addition, this research relied on resources provided to the eBOSS Collaboration by the National Energy Research Scientific Computing Center (NERSC). NERSC is a U.S.\ Department of Energy Office of Science User Facility operated under Contract No.\ DE-AC02-05CH11231.

SDSS-IV is managed by the Astrophysical Research Consortium for the Participating Institutions of the SDSS Collaboration including the Brazilian Participation Group, the Carnegie Institution for Science, Carnegie Mellon University, the Chilean Participation Group, the French Participation Group, Harvard-Smithsonian Center for Astrophysics, Instituto de Astrof{\'i}sica de Canarias, The Johns Hopkins University, Kavli Institute for the Physics and Mathematics of the Universe (IPMU), University of Tokyo, the Korean Participation Group, Lawrence Berkeley National Laboratory, Leibniz Institut f{\"u}r Astrophysik Potsdam (AIP), Max-Planck-Institut f{\"u}r Astronomie (MPIA Heidelberg), Max-Planck-Institut f{\"u}r Astrophysik (MPA Garching), Max-Planck-Institut f{\"u}r Extraterrestrische Physik (MPE), National Astronomical Observatories of China, New Mexico State University, New York University, University of Notre Dame, Observat{\'a}rio Nacional, MCTI, The Ohio State University, Pennsylvania State University, Shanghai Astronomical Observatory, United Kingdom Participation Group, Universidad Nacional Aut{\'o}noma de M{\'e}xico, University of Arizona, University of Colorado Boulder, University of Oxford, University of Portsmouth, University of Utah, University of Virginia, University of Washington, University of Wisconsin, Vanderbilt University, and Yale University.

\bibliography{allqsos}

\clearpage
\appendix
\section{Statistical uncertainties in pipeline and PCA redshifts}\label{asec:zstatuncertainties}
\begin{figure*}[ht]
    \epsscale{0.6}
    \plotone{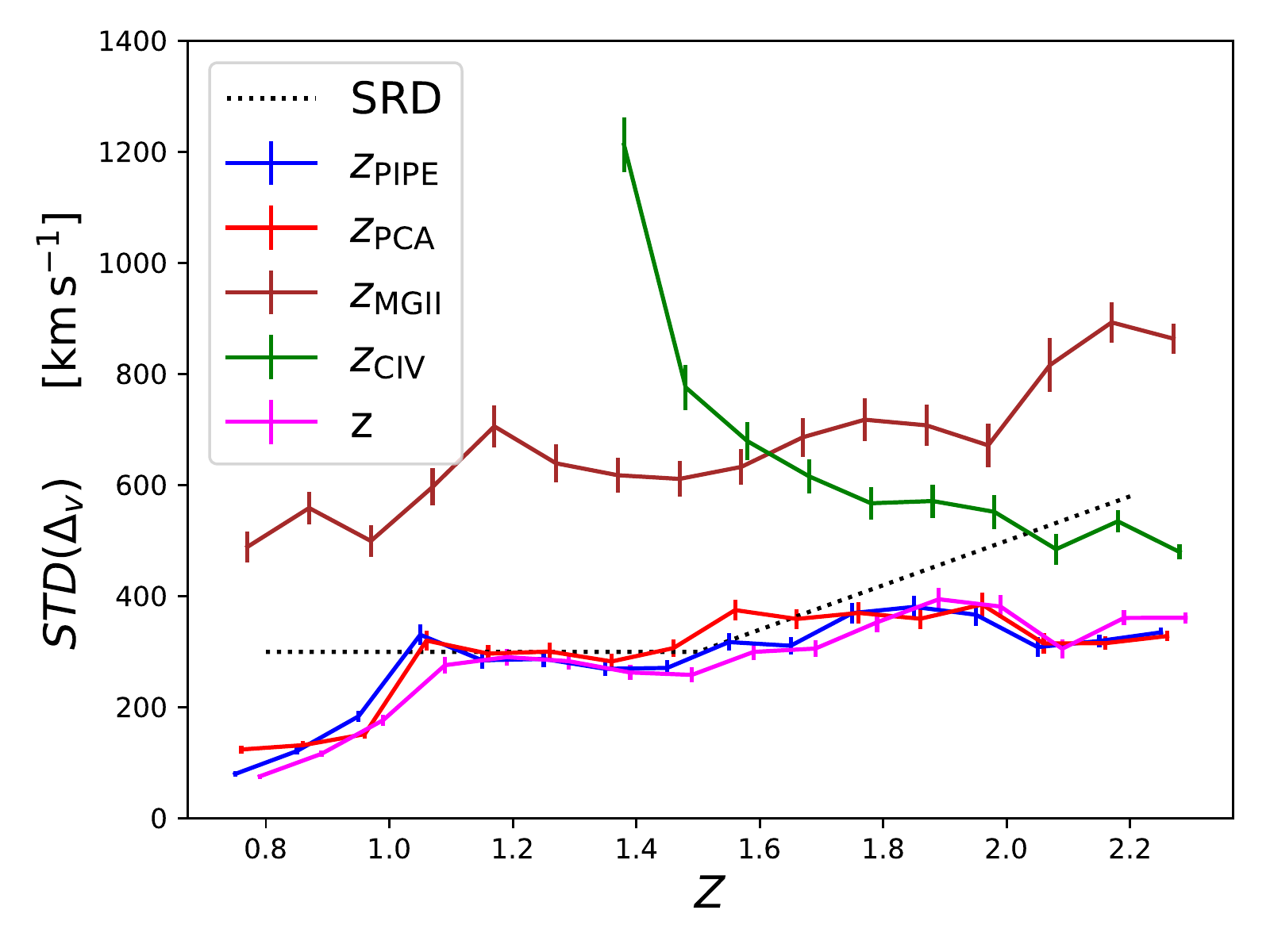}
    \caption{The distribution of statistical uncertainty for the redshift sources appearing in Fig.~\ref{fig:z_v_sys} modeled with the redshift-dependent double Gaussian profile defined in Eqn.~\ref{eqn:doublegauss}. {\it Z} (pink solid line) meets the requirement for statistical uncertainty defined in \citet[][black dotted line]{eboss_mission}.\label{fig:sigmalines}}
\end{figure*}
Fig.~\ref{fig:sigmalines} shows the distribution of the statistical uncertainty for \zpipe\ and \zpca\ together with a redshift-dependent Gaussian distribution following the eBOSS requirements listed in \citet{eboss_mission} for ``low'' ($z < 1.5$) and ``high'' redshift ($z > 1.5$) quasars\footnote{Specifically: \S3.2, requirement 1, of \citet{eboss_mission}.}. As noted in \S\ref{ssec:redshiftcomparisons}, the actual distribution of redshift uncertainty contains tails extending to $|\Delta v| = 3000$\,\kms, which are not captured by a single Gaussian model. Therefore, we considered a more realistic model based on a double-Gaussian profile (see the black curve in Fig.~\ref{fig:deltav_sys}) defined by:
\begin{equation}\label{eqn:doublegauss}
    \frac{dN}{d(\delta \nu)}=\frac{1}{\sqrt{2\pi(1+f)}}
    \left[
    \frac{f}{\sigma_1}\exp{\frac{-\Delta\nu^2}{2\sigma_1^2}}
    \frac{1}{\sigma_2}\exp{\frac{-\Delta\nu^2}{2\sigma_2^2}}
    \right],
\end{equation}
where both Gaussian functions are centered on zero. The resulting fit to the data gives $\sigma_1 = 150$\,\kms, $\sigma_2 = 1000$\,\kms, and $f = 4.478$. We kept the shape of this distribution fixed with redshift. A detailed study of the impact of this modeled, double-Gaussian redshift uncertainty on the cosmological parameters, using N-body mock catalogs, is presented in \citet{smith20}. In addition, the clustering analyses presented in \citet{hou20a} incorporated this redshift uncertainty when deriving cosmological parameters. Based on the Hou et al.\ and Smith et al.\ analyses, the total errors (statistical and systematic) of \zpipe\ and \zpca\ meet the eBOSS requirements detailed in \citet{eboss_mission}, provided that we model the shape of the redshift accuracy and precision using Eqn.~\ref{eqn:doublegauss}.

As detailed in \S\ref{ssec:redshiftcomparisons}, we found that the catastrophic failure rate for DR16Q quasar redshifts is of order 1\%, meeting the requirements quoted in \citet{eboss_mission}. In \citet{smith20}, the effect of the catastrophic failures was implemented in N-body mock catalogs, and the impact of these is studied on cosmological constraints. They found that even a catastrophic failure rate of over 1\% is sufficiently small that constraints on the cosmological parameters from eBOSS quasars are unaffected.

\section{Interesting spectra}\label{asec:interestingspectra}
DR16Q includes a wide range of illustrative quasar spectra. We have chosen to include four such interesting, unusual, or informative spectra in this appendix. For all of the spectra plotted here, the line rest wavelengths were taken from Table~2 of \citet{vandenberk2001}.

\begin{figure*}[ht]
    \epsscale{1.0}
    \plotone{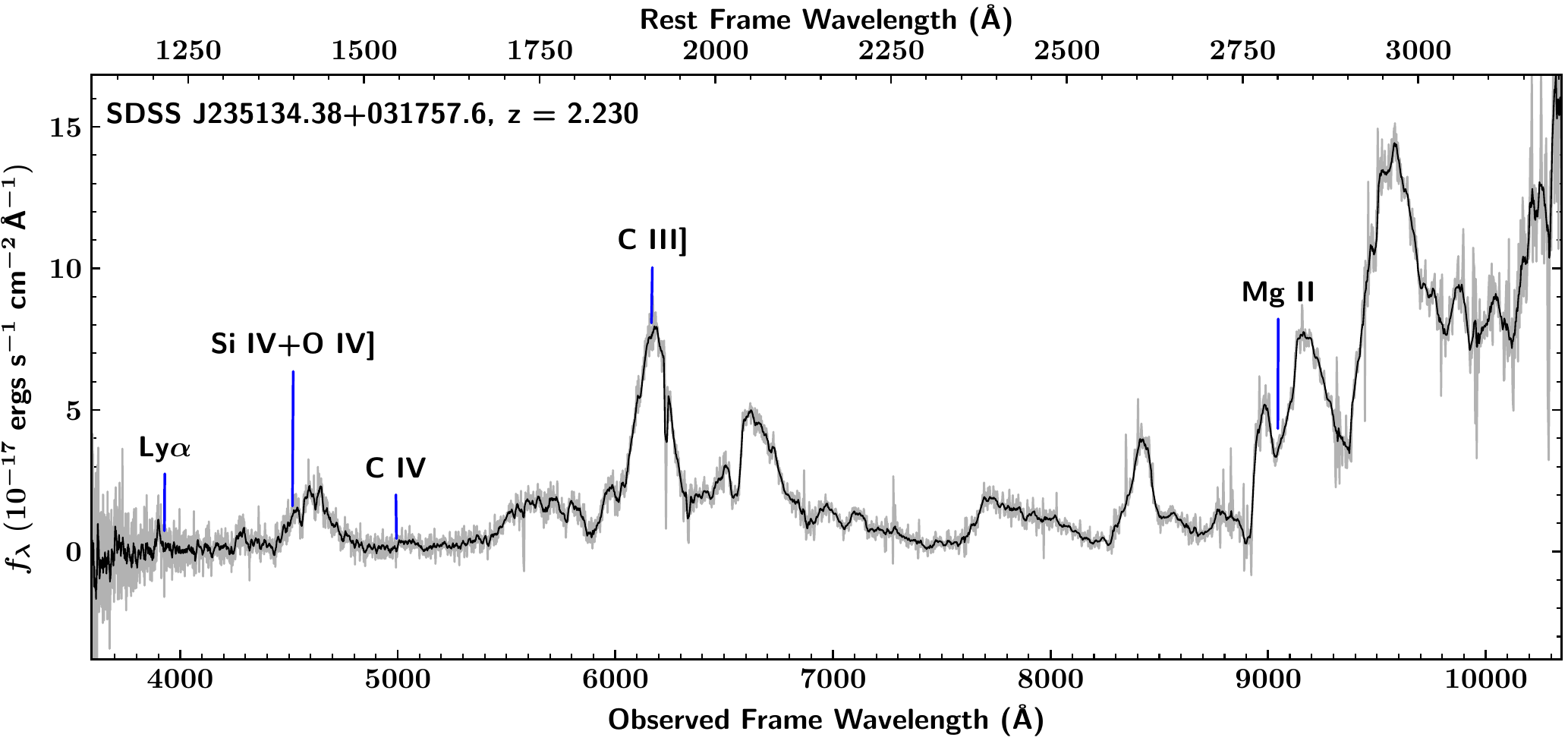}
    \caption{The FeLoBAL quasar SDSS\,J235134.38+031757.6, included in DR16Q under plate number 8741, MJD 57390, and fiber ID 450. Rest-frame wavelengths are plotted using the redshift $z = 2.230$ (obtained via a cross-correlation with similar FeLoBALs), although this spectrum was randomly selected for visual inspection (see \S\ref{ssec:randomvi}) and assigned ${\tt Z\_VI} = 2.43$. Locations of common quasar broad emission lines have been labeled (based on $z = 2.230$), though they do not obviously align with the broad features.\label{fig:spectest0}}
\end{figure*}

In Fig.~\ref{fig:spectest0}, we present the spectrum of an Iron Low-Ionization Broad Absorption Line (FeLoBAL) quasar discovered during the random visual inspections. For more information on these unusual objects see \S5.2 and Fig.~6 of \citet{hall2002}.

\begin{figure*}[ht]
    \epsscale{1.0}
    \plotone{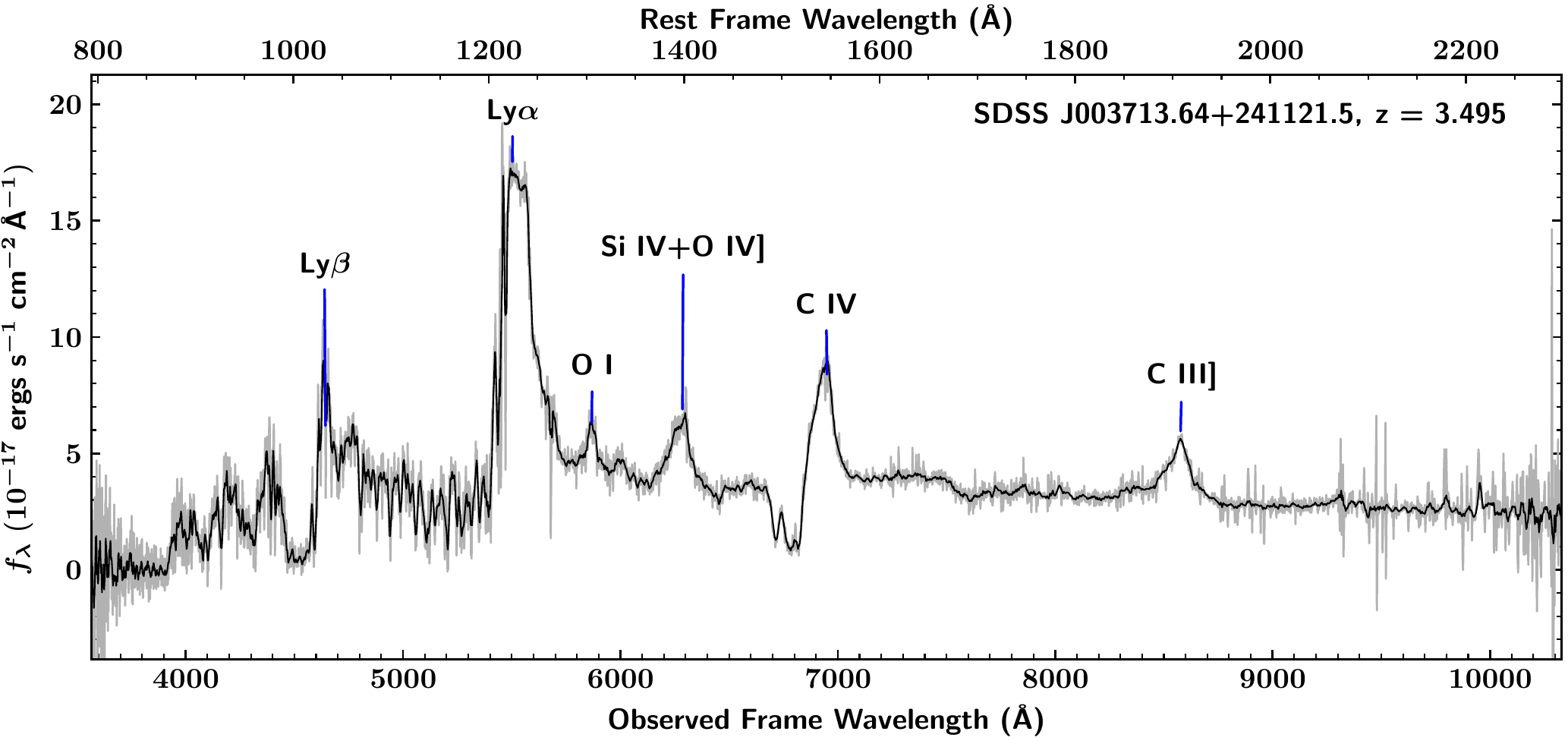}
    \caption{Quasar SDSS\,J003713.64+241121.5, included in DR16Q under plate number 7672, MJD 57339, and fiber ID 394, and with a redshift of $\texttt{Z\_10K} = 3.495$. This spectrum is an example of a BAL quasar. The Lyman limit can be clearly seen at ~$3900$\,\AA\ (observed frame). The spectrum also displays a typically weak \ion{O}{1}\,$\lambda1305$ broad emission line, which usually has flux relative to \lya\ of $\sim2$\%~\citep{vandenberk2001}. An (unlabeled) weak \ion{C}{2}\,$\lambda1336$ emission line can also be seen, which \citet{vandenberk2001} reports having a flux relative to \lya\ of $\sim0.7$\%.
    \label{fig:spectest1}}
\end{figure*}

DR16Q includes many BAL quasars. In Fig.~\ref{fig:spectest1} we present a quasar that shows BAL features. These BAL features were identified by the algorithm described in \S\ref{ssec:bals}. The spectrum also displays two emission lines that are often weak in quasars, \ion{O}{1}\,$\lambda1305$ (labeled) and \ion{C}{2}\,$\lambda1336$ (unlabeled).

\begin{figure*}[ht]
    \epsscale{1.0}
    \plotone{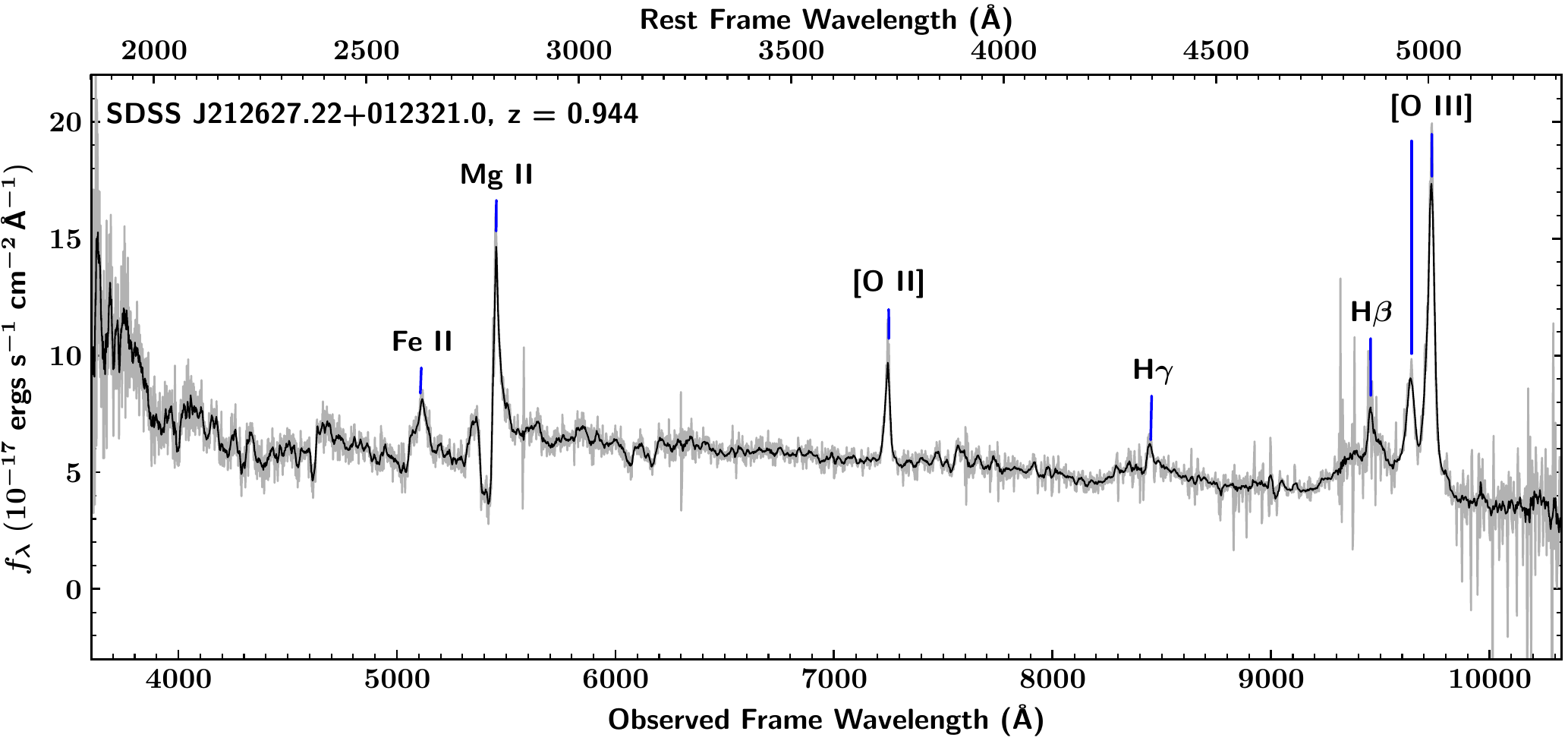}
    \caption{Quasar SDSS\,J212627.22+012321.0, included in DR16Q under plate number 9162, MJD 58040, fiber ID 354, and with a redshift of $\texttt{Z\_10K} = 0.944$. The spectrum shows a clear \mgii\ BAL feature along with a broad \ion{Fe}{2}\,$\lambda2627$ emission line.
    \label{fig:spectest2}}
\end{figure*}

The BAL algorithm described in \S\ref{ssec:bals} focuses on \civ\ and \siiv\ features. In Fig.~\ref{fig:spectest2} we present a different type: an \mgii\ BAL. This spectrum was {\em not} identified by the BAL algorithm as the quasar is not at a high enough redshift for \civ\ or \siiv\ to be in the SDSS observable wavelength range.

\begin{figure*}[ht]
    \epsscale{1.0}
    \plotone{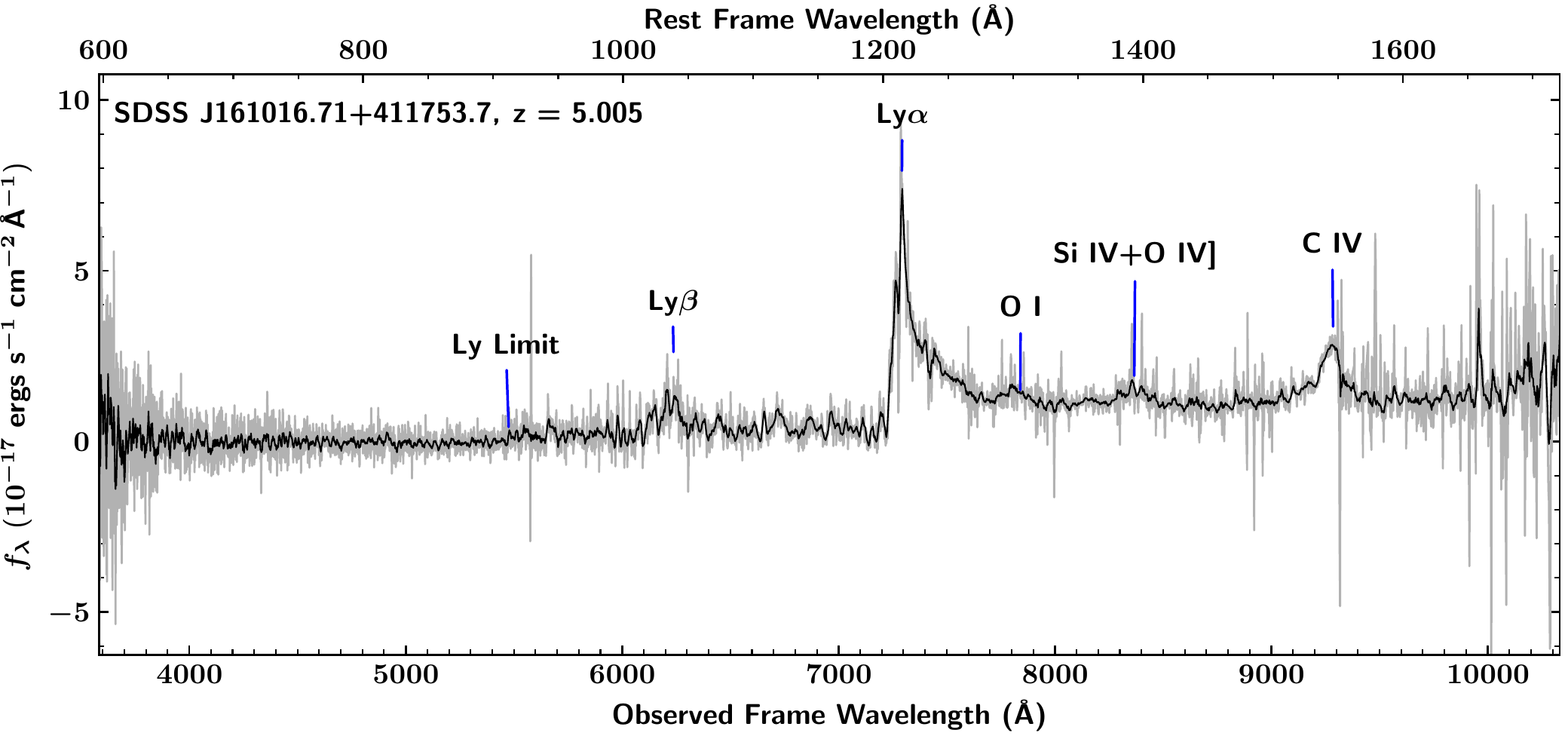}
    \caption{Quasar SDSS\,J161016.71+411753.7, included in DR16Q under plate number 6044, MJD 56090, fiber ID 418, and with a redshift of $\texttt{Z\_VI} = 5.005$. This higher-redshift spectrum is included as an example of a strong \lya\ emission line with a visible, but weak \lya\ forest. The (weak) Lyman limit feature is labeled to show the edge of the forest region. The same \ion{O}{1}\,$\lambda1305$ feature seen in Fig.~\ref{fig:spectest1} can be seen in this spectrum, though at a significantly reduced strength.\label{fig:spectest3}}
\end{figure*}

SDSS-III/BOSS and SDSS-IV/eBOSS focused, in part, on studies of the \lya\ forest using $z > 2.2$ quasars. In Fig.~\ref{fig:spectest3} we include a spectrum of one such target that has a clearly visible \lya\ forest. This spectrum highlights some rarer emission features and a weak Lyman limit.

\clearpage
\section{Issues from Previous Quasar Catalogs}\label{asec:oldbugs}
As large scale usage of the SDSS quasar catalogs has increased over time, bugs have been discovered within previous iterations of these catalogs. Documented bugs and other issues that appeared in the twelfth and fourteenth catalogs have been corrected for this iteration of the quasar-only and superset catalogs. We list the known issues below. All issues listed have been corrected in the most recent version of the catalogs released: DR16Q\_v4 and DR16Q\_Superset\_v3 with two exceptions explained below.
\begin{enumerate}
    \item (DR12Q) Incorrect astrometry for DR12Q objects.
    \begin{enumerate}
        \item DR12Q-added objects now use the current astrometric coordinates for BOSS objects.
    \end{enumerate}
    \item (DR12Q) Absolute {\it i}-band magnitudes were incorrectly calculated. K-corrections were only done to the continuum and did not include emission line corrections. Additionally it appears the set cosmology parameters used was not either of the two sets published in the paper.
    \begin{enumerate}
        \item K-correction values were correctly applied and cosmology parameters match those listed in DR16Q.
    \end{enumerate}
    \item (DR12Q, DR14Q) Spectra with \texttt{ZWARNING} flags for \texttt{UNPLUGGED} or \texttt{BAD\_TARGET} were included.
    \begin{enumerate}
        \item All records with these \texttt{ZWARNING} flags have been removed.
    \end{enumerate}
    \item (DR14Q) Opening DR14Q reports a ``VOTable opening error.''
    \begin{enumerate}
        \item VOTable errors have been corrected and the first table HDU in the FITS files have been named ``CATALOG.''
    \end{enumerate}
    \item (DR14Q) \texttt{DUPLICATE} fields are populated with errant -1 values at every other position.
    \begin{enumerate}
        \item \texttt{DUPLICATE} fields have been populated with the correct \texttt{PLATE}, \texttt{MJD}, \texttt{FIBERID}, and \texttt{SPECTRO} values.
    \end{enumerate}
    \item (DR14Q) SDSS names incorrectly reported due to erroneous rounding errors.
    \begin{enumerate}
        \item SDSS names are now correctly truncated, rather than rounded.
    \end{enumerate}
    \item (DR14Q) \texttt{OBJ\_ID} fields are blank.
    \begin{enumerate}
        \item The column has been renamed to \texttt{OBJID} to match other SDSS catalogs, and has been populated for BOSS and eBOSS objects. \texttt{OBJID} values for SDSS-I/II quasars have been included where they were available.
    \end{enumerate}
    \item (DR14Q) \texttt{GALEX\_MATCHED} and \texttt{UKIDSS\_MATCHED} are recorded as floating numbers instead of integers.
    \begin{enumerate}
        \item These two fields are now recorded as integers, and use 0 or 1 values to denote no-match or a match respectively.
    \end{enumerate}
    \item (DR14Q) \texttt{FIRST\_MATCHED} uses incorrect values.
    \begin{enumerate}
        \item The \texttt{FIRST\_MATCHED} column now uses 0 or 1 to denote a match.
    \end{enumerate}
    \item (DR14Q) Records with no XMM or 2MASS match are blank.
    \begin{enumerate}
        \item Quasars with no match in XMM or 2MASS now have -1 values rather than blank entries.
    \end{enumerate}
    \item (DR14Q) The following columns all had zero values or were empty: \texttt{RUN\_NUMBER}, \texttt{COL\_NUMBER}, \texttt{RERUN\_NUMBER}, \texttt{FIELD\_NUMBER}, \texttt{SPECTRO\_DUPLICATE}.
    \begin{enumerate}
        \item These columns have been correctly populated.
    \end{enumerate}
    \item (DR14Q) \texttt{N\_SPEC} values did not match the number of additional spectra due to incorrect population of -1 values in the \texttt{DUPLICATE} fields (see \#5 above).
    \begin{enumerate}
        \item The \texttt{NSPEC} fields have been populated correctly and renamed.
    \end{enumerate}
    \item (DR14Q) Hyphens and periods appear in some column names. These may cause FITS reader programs to report errors.
    \begin{enumerate}
        \item All column names with hyphens or periods have been renamed to remove these punctuation marks. All columns now only use alphanumeric characters and underscores.
    \end{enumerate}
    \item (DR14Q) Some TUNITS FITS header cards do not match the reported units.
    \begin{enumerate}
        \item All TUNITS header cards have been removed. Refer to this paper or the datamodel\footnote{\url{https://data.sdss.org/datamodel/files/BOSS\_QSO/DR16Q/}} for the correct units.
    \end{enumerate}
    \item (DR14Q) The \texttt{Z\_ERR} column did not contain errors for pipeline redshifts as reported.
    \begin{enumerate}
        \item The \texttt{Z\_ERR} column has been removed.
    \end{enumerate}
    \item (DR14Q) The \texttt{ZWARNING} column was not populated.
    \begin{enumerate}
        \item The \texttt{ZWARNING} column has been correctly populated for BOSS and eBOSS quasars.
    \end{enumerate}
    \item (DR16Q) Six known quasars from DR7Q were not added to DR16Q and could not be included without affecting other SDSS projects in progress. These quasars can be found in the Science Archive Server (SAS) under the following PLATE-MJD-FIBERID combinations: 901-52641-307, 1194-52703-58, 1611-53147-507, 1768-53442-57, 1948-53388-42, 2784-54529-73.
\end{enumerate}

\clearpage
\section{FITS Table Description}\label{asec:append}
\startlongtable
\begin{deluxetable*}{clcl}
\tabletypesize{\footnotesize}
\tablecaption{Format of the FITS binary table containing DR16Q \label{tab:fields}}
\tablecolumns{4}
\tablewidth{0pt}
\tablehead{
\colhead{Column} & \colhead{Name} & \colhead{Format} & \colhead{Description}
}
\startdata
1 & SDSS\_NAME & STRING & SDSS DR16 designation - hh:mm:ss.ss$\pm$dd:mm:ss.s (J2000)\\
2 & RA & DOUBLE & Right Ascension in decimal degrees (J2000)\\
3 & DEC & DOUBLE & Declination in decimal degrees (J2000)\\
4 & PLATE & INT32 & Spectroscopic plate number\\
5 & MJD & INT32 & Modified Julian day of the spectroscopic observation \\
6 & FIBERID & INT16 & Spectroscopic fiber number \\
\hline
7 & AUTOCLASS\_PQN & STRING & Object classification post-\qnet\ \\
8 & AUTOCLASS\_DR14Q & STRING & Object classification based only on the DR14Q algorithm \\
9 & IS\_QSO\_QN & INT16 & Binary flag for \qnet\ quasar identification \\
10 & Z\_QN & DOUBLE & Redshift derived by \qnet\ \\
\hline
11 & RANDOM\_SELECT & INT16 & Binary flag indicating objects selected for random visual inspection \\
12 & Z\_10K & DOUBLE & Redshift from visual inspection of random set \\
13 & Z\_CONF\_10K & INT16 & Confidence rating for visual inspection redshift of random set \\
14 & PIPE\_CORR\_10K & INT16 & Binary flag indicating if the automated pipeline classification \\
 & & & and redshift were correct in the random set \\
15 & IS\_QSO\_10K & INT16 & Binary flag for random set quasar identification \\
16 & THING\_ID & INT64 & SDSS identifier \\
\hline
17 & Z\_VI & DOUBLE & Redshift from visual inspection \\
18 & Z\_CONF & INT16 & Confidence rating for visual inspection redshift \\
19 & CLASS\_PERSON & INT32 & Object classification from visual inspection \\
20 & Z\_DR12Q & DOUBLE & Redshift taken from DR12Q visual inspection \\
21 & IS\_QSO\_DR12Q & INT16 & Flag indicating if an object was a quasar in DR12Q \\
22 & Z\_DR7Q\_SCH & DOUBLE & Redshift taken from DR7Q \citet{dr7q_paper} catalog \\
23 & IS\_QSO\_DR7Q & INT16 & Flag indicating if an object was a quasar in DR7Q \\
24 & Z\_DR6Q\_HW & DOUBLE & Redshift taken from DR6 \citet{dr6q_hw_paper} catalog \\
25 & Z\_DR7Q\_HW & DOUBLE & Redshift taken from the \citet{shen2011} catalog ancillary columns.\footnote{See column 143 at \url{http://das.sdss.org/va/qso_properties_dr7/dr7.htm}} \\
26 & IS\_QSO\_FINAL & INT16 & Flag indicating quasars included in final catalog \\
27 & Z & DOUBLE & Best available redshift taken from Z\_VI, Z\_PIPE, Z\_DR12Q, \\
 & & & Z\_DR7Q\_SCH, or Z\_DR6Q\_HW \\
28 & SOURCE\_Z & STRING & Source for redshift in previous column \\
29 & Z\_PIPE & DOUBLE & Redshift from SDSS pipeline \\
30 & ZWARNING & INT32 & Quality flag on SDSS pipeline redshift estimate \\
31 & OBJID & STRING & SDSS object identification number \\
\hline
32 & Z\_PCA & DOUBLE & PCA redshift derived by the \rvb\ algorithm \\
33 & ZWARN\_PCA & INT64 & Warning flag for \rvb\ redshift \\
34 & DELTACHI2\_PCA & DOUBLE & $\Delta \chi^{2}$ for PCA redshift vs. cubic continuum fit \\
35 & Z\_HALPHA & DOUBLE & PCA line redshift for \ha\ from \rvb\ \\
36 & ZWARN\_HALPHA & INT64 & Warning flag for \ha \\
37 & DELTACHI2\_HALPHA & DOUBLE & $\Delta \chi^{2}$ for \ha\ line redshift vs. cubic continuum fit \\
38 & Z\_HBETA & DOUBLE & PCA line redshift for \hb\ from \rvb\ \\
39 & ZWARN\_HBETA & INT64 & Warning flag for \hb \\
40 & DELTACHI2\_HBETA & DOUBLE & $\Delta \chi^{2}$ for \hb\ line redshift vs. cubic continuum fit \\
41 & Z\_MGII & DOUBLE & PCA line redshift for \mgii\ from \rvb\ \\
42 & ZWARN\_MGII & INT64 & Warning flag for \mgii \\
43 & DELTACHI2\_MGII & DOUBLE & $\Delta \chi^{2}$ for \mgii\ line redshift vs. cubic continuum fit \\
44 & Z\_CIII & DOUBLE & PCA line redshift for \ciii\ from \rvb\ \\
45 & ZWARN\_CIII & INT64 & Warning flag for \ciii \\
46 & DELTACHI2\_CIII & DOUBLE & $\Delta \chi^{2}$ for \ciii\ line redshift vs. cubic continuum fit \\
47 & Z\_CIV & DOUBLE & PCA line redshift for \civ\ from \rvb\ \\
48 & ZWARN\_CIV & INT64 & Warning flag for \civ \\
49 & DELTACHI2\_CIV & DOUBLE & $\Delta \chi^{2}$ for \civ\ line redshift vs. cubic continuum fit \\
50 & Z\_LYA & DOUBLE & PCA line redshift for \lya\ from \rvb\ \\
51 & ZWARN\_LYA & INT64 & Warning flag for \lya \\
52 & DELTACHI2\_LYA & DOUBLE & $\Delta \chi^{2}$ for \lya\ line redshift vs. cubic continuum fit \\
53 & Z\_LYAWG & FLOAT & PCA redshift derived from spectra masked blueward of \lya.\\
\hline
54 & Z\_DLA & DOUBLE[5] & Redshift for damped \lya\ features \\
55 & NHI\_DLA & DOUBLE[5] & Absorber column density for damped \lya\ features \\
56 & CONF\_DLA & DOUBLE[5] & Confidence of detection for damped \lya\ features \\
57 & BAL\_PROB & FLOAT & BAL probability \\
58 & BI\_CIV & DOUBLE & BALnicity index for \civ\ region in \kms\\
59 & ERR\_BI\_CIV & DOUBLE & Uncertainty of BI for \civ\ region in \kms\\
60 & AI\_CIV & DOUBLE & Absorption index for \civ\ region in \kms\\
61 & ERR\_AI\_CIV & DOUBLE & Uncertainty of absorption index for \civ\ region in \kms\\
62 & BI\_SIIV & DOUBLE & BALnicity index for \siiv\ region in \kms\\
63 & ERR\_BI\_SIIV & DOUBLE & Uncertainty of BI for \siiv\ region in \kms\\
64 & AI\_SIIV & DOUBLE & Absorption index for \siiv\ region in \kms\\
65 & ERR\_AI\_SIIV & DOUBLE & Uncertainty of absorption index for \siiv\ region in \kms\\
\hline
66 & BOSS\_TARGET1 & INT64 & BOSS target selection flag for main survey \\
67 & EBOSS\_TARGET0 & INT64 & eBOSS target selection flag for pilot survey (SEQUELS) \\
68 & EBOSS\_TARGET1 & INT64 & eBOSS target selection flag for main survey \\
69 & EBOSS\_TARGET2 & INT64 & eBOSS target selection flag for main survey \\
70 & ANCILLARY\_TARGET1 & INT64 & BOSS target selection flag for ancillary surveys \\
71 & ANCILLARY\_TARGET2 & INT64 & BOSS target selection flag for ancillary surveys \\
\hline
72 & NSPEC\_SDSS & INT32 & Number of additional spectra from SDSS-I/II \\
73 & NSPEC\_BOSS & INT32 & Number of additional spectra from BOSS/eBOSS \\
74 & NSPEC & INT32 & Total Number of additional spectra \\
75 & PLATE\_DUPLICATE & INT32[74] & Spectroscopic plate number for duplicate spectrum \\
76 & MJD\_DUPLICATE & INT32[74] & Spectroscopic MJD for duplicate spectrum \\
77 & FIBERID\_DUPLICATE & INT16[74] & Spectroscopic fiber number for duplicate spectrum \\
78 & SPECTRO\_DUPLICATE & INT32[74] & Spectroscopic epoch for each duplicate, 1=SDSS, 2=(e)BOSS \\
\hline
79 & SKYVERSION & INT8 & SDSS photometric sky version number \\
80 & RUN\_NUMBER & INT32 & SDSS imaging run number of photometric measurements \\
81 & RERUN\_NUMBER & STRING & SDSS photometric processing rerun number \\
82 & CAMCOL\_NUMBER & INT32 & SDSS camera column number (1--6) \\
83 & FIELD\_NUMBER & INT32 & SDSS field number \\
84 & ID\_NUMBER & INT32 & SDSS photometric ID number \\
85 & LAMBDA\_EFF & DOUBLE & Wavelength to optimize hold location for, in \AA \\
86 & ZOFFSET & DOUBLE & Backstopping offset distance in \micm \\
87 & XFOCAL & DOUBLE & Hole x-axis position in focal plane in \mm \\
88 & YFOCAL & DOUBLE & Hole y-axis position in focal plane in \mm \\
89 & CHUNK & STRING & Name of tiling chunk (from platelist product) \\
90 & TILE & INT32 & Tile number \\
91 & PLATESN2 & DOUBLE & Overall (S/N)$^{2}$ measure for plate, minimum of all four cameras \\
\hline
92 & PSFFLUX & FLOAT[5] & Flux in \textit{u}, \textit{g}, \textit{r}, \textit{i}, \textit{z} bands \\
93 & PSFFLUX\_IVAR & FLOAT[5] & Inverse variance of the \textit{u}, \textit{g}, \textit{r}, \textit{i}, \textit{z} fluxes \\
94 & PSFMAG & FLOAT[5] & PSF magnitudes in \textit{u}, \textit{g}, \textit{r}, \textit{i}, \textit{z} bands \\
95 & PSFMAGERR & FLOAT[5] & Error on PSF magnitudes in \textit{u}, \textit{g}, \textit{r}, \textit{i}, \textit{z} bands \\
96 & EXTINCTION & FLOAT[5] & Galactic extinction in the five SDSS bands \citep{schlafly2011} \\
97 & M\_I & FLOAT & Absolute \textit{i}-band magnitude, $M_i$[z=2] using \texttt{Z\_PCA} redshift in Col.\ 32\\
 & & & See \S\ref{sec:description} for details on how $M_{i}$[$z=2$] was calculated.\\
98 & SN\_MEDIAN\_ALL & DOUBLE & S/N median value for all good spectroscopic pixels \\
\hline
99 & GALEX\_MATCHED & INT16 & GALEX matching flag \\
100 & FUV & DOUBLE & \textit{fuv} flux for GALEX \\
101 & FUV\_IVAR & DOUBLE & Inverse variance of \textit{fuv} flux\\
102 & NUV & DOUBLE & \textit{nuv} flux for GALEX \\
103 & NUV\_IVAR & DOUBLE & Inverse variance of \textit{nuv} flux\\
\hline
104 & UKIDSS\_MATCHED & INT16 & UKIDSS matching flag\\
105 & YFLUX & DOUBLE & {\it Y}-band flux density from UKIDSS (in~\wmh) \\
106 & YFLUX\_ERR & DOUBLE & Error in {\it Y}-band flux density from UKIDSS (in~\wmh) \\
107 & JFLUX & DOUBLE & {\it J}-band flux density from UKIDSS (in~\wmh) \\
108 & JFLUX\_ERR & DOUBLE & Error in {\it J}-band flux density from UKIDSS (in~\wmh) \\
109 & HFLUX & DOUBLE & {\it H}-band flux density from UKIDSS (in~\wmh) \\
110 & HFLUX\_ERR & DOUBLE & Error in {\it H}-band flux density from UKIDSS (in~\wmh) \\
111 & KFLUX & DOUBLE & {\it K}-band flux density from UKIDSS (in~\wmh) \\
112 & KFLUX\_ERR & DOUBLE & Error in {\it K}-band flux density from UKIDSS (in~\wmh) \\
\hline
113 & W1\_FLUX & FLOAT & WISE flux in W1-band ($3.4$\,\micm) (Vega, nanomaggies) \\
114 & W1\_FLUX\_IVAR & FLOAT & Inverse variance of the W1-band flux (Vega, nanomaggies$^{-2}$) \\
115 & W1\_MAG & FLOAT & W1-band magnitude (Vega) \\
116 & W1\_MAG\_ERR & FLOAT & W1-band uncertainty in magnitude (Vega) \\
117 & W1\_CHI2 & FLOAT & Profile-weighted $\chi^{2}$ for W1-band \\
118 & W1\_FLUX\_SNR & FLOAT & Signal-to-Noise ratio for W1-band from flux and inverse variance \\
119 & W1\_SRC\_FRAC & FLOAT & Fraction of source in profile-weighted fit \\
120 & W1\_EXT\_FLUX & FLOAT & Amount of flux added to the profile by nearby sources \\
121 & W1\_EXT\_FRAC & FLOAT & Fraction of profile flux from nearby sources \\
122 & W1\_NPIX & INT16 & Number of pixels included in profile fit \\
123 & W2\_FLUX & FLOAT & WISE flux in W2-band ($4.6$\,\micm) (Vega, nanomaggies) \\
124 & W2\_FLUX\_IVAR & FLOAT & Inverse variance of the W2-band flux (Vega, nanomaggies$^{-2}$) \\
125 & W2\_MAG & FLOAT & W2-band magnitude (Vega) \\
126 & W2\_MAG\_ERR & FLOAT & W2-band uncertainty in magnitude (Vega) \\
127 & W2\_CHI2 & FLOAT & Profile-weighted $\chi^{2}$ for W2-band \\
128 & W2\_FLUX\_SNR & FLOAT & Signal-to-Noise ratio for W2-band from flux and inverse variance \\
129 & W2\_SRC\_FRAC & FLOAT & Fraction of source in profile-weighted fit \\
130 & W2\_EXT\_FLUX & FLOAT & Amount of flux added to the profile by nearby sources \\
131 & W2\_EXT\_FRAC & FLOAT & Fraction of profile flux from nearby sources \\
132 & W2\_NPIX & INT16 & Number of pixels included in profile fit \\
\hline
133 & FIRST\_MATCHED & INT16 & Matching flag for FIRST \\
134 & FIRST\_FLUX & DOUBLE & FIRST peak flux density at 20\,\cm\ in \mjy \\
135 & FIRST\_SNR & DOUBLE & FIRST flux density S/N \\
136 & SDSS2FIRST\_SEP & DOUBLE & SDSS-FIRST separation in arcsec \\
\hline
137 & JMAG & DOUBLE & 2MASS {\it J}-band magnitude (Vega) \\
138 & JMAG\_ERR & DOUBLE & 2MASS error in {\it J}-band magnitude \\
139 & JSNR & DOUBLE & 2MASS {\it J}-band S/N \\
140 & JRDFLAG & INT32 & 2MASS {\it J}-band photometry flag \\
141 & HMAG & DOUBLE & 2MASS {\it H}-band magnitude (Vega) \\
142 & HMAG\_ERR & DOUBLE & 2MASS error in {\it H}-band magnitude \\
143 & HSNR & DOUBLE & 2MASS {\it H}-band S/N \\
144 & HRDFLAG & INT32 & 2MASS {\it H}-band photometry flag \\
145 & KMAG & DOUBLE & 2MASS $K_s$-band magnitude (Vega) \\
146 & KMAG\_ERR & DOUBLE & 2MASS error in $K_s$-band magnitude \\
147 & KSNR & DOUBLE & 2MASS $K_s$-band S/N \\
148 & KRDFLAG & INT32 & 2MASS $K_s$-band photometry flag \\
149 & SDSS2MASS\_SEP & DOUBLE & SDSS-2MASS separation in arcsec \\
\hline
150 & 2RXS\_ID & STRING & ROSAT ID \\
151 & 2RXS\_RA & DOUBLE & Right ascension of the ROSAT source in decimal degrees (J2000) \\
152 & 2RXS\_DEC & DOUBLE & Declination of the ROSAT source in decimal degrees (J2000) \\
153 & 2RXS\_SRC\_FLUX & FLOAT & ROSAT source flux in 0.5--2.0\,\kev\ band in \ecs \\
154 & 2RXS\_SRC\_FLUX\_ERR & FLOAT & ROSAT source flux error in 0.5--2.0\,\kev\ band in \ecs \\
155 & SDSS2ROSAT\_SEP & DOUBLE & SDSS-ROSAT separation in arcsec \\
\hline
156 & XMM\_SRC\_ID & INT64 & XMM source ID \\
157 & XMM\_RA & DOUBLE & Right ascension for XMM source in decimal degrees (J2000) \\
158 & XMM\_DEC & DOUBLE & Declination for XMM source in decimal degrees (J2000) \\
159 & XMM\_SOFT\_FLUX & FLOAT & Soft (0.2--2.0\,\kev) X-ray flux in \ecs \\
160 & XMM\_SOFT\_FLUX\_ERR & FLOAT & Error on soft X-ray flux in \ecs \\
161 & XMM\_HARD\_FLUX & FLOAT & Hard (2.0--12.0\,\kev) X-ray flux in \ecs \\
162 & XMM\_HARD\_FLUX\_ERR & FLOAT & Error on hard X-ray flux from in \ecs \\
163 & XMM\_TOTAL\_FLUX & FLOAT & Total (0.2--12.0\,\kev) X-ray flux in \ecs \\
164 & XMM\_TOTAL\_FLUX\_ERR & FLOAT & Error on total X-ray flux in \ecs \\
165 & XMM\_TOTAL\_LUM & FLOAT & Total (0.2--12.0\,\kev) X-ray luminosity in \esec\\
166 & SDSS2XMM\_SEP & DOUBLE & SDSS-XMM-Newton separation in arcsec \\
\hline
167 & GAIA\_MATCHED & INT16 & Gaia matching flag\\
168 & GAIA\_DESIGNATION & STRING & Gaia designation, includes data release and source ID in that release\\
169 & GAIA\_RA & DOUBLE & Gaia barycentric right ascension in decimal degrees (J2015.5)\\
170 & GAIA\_DEC & DOUBLE & Gaia barycentric right declination in decimal degrees (J2015.5)\\
171 & GAIA\_PARALLAX & DOUBLE & Absolute stellar parallax (J2015.5) \\
172 & GAIA\_PARALLAX\_ERR & DOUBLE & Standard error of the stellar parallax (J2015.5)\\
173 & GAIA\_PM\_RA & DOUBLE & Proper motion in right ascension (\masy, J2015.5)\\
174 & GAIA\_PM\_RA\_ERR & DOUBLE & Standard error of the proper motion in RA (\masy, J2015.5)\\
175 & GAIA\_PM\_DEC & DOUBLE & Proper motion in declination (\masy, J2015.5)\\
176 & GAIA\_PM\_DEC\_ERR & DOUBLE & Standard error of the proper motion in DEC (\masy, J2015.5)\\
177 & GAIA\_G\_MAG & DOUBLE & Mean magnitude in G-band (Vega) \\
178 & GAIA\_G\_FLUX\_SNR & DOUBLE & Mean flux over standard deviation in G-band (Vega) \\
179 & GAIA\_BP\_MAG & DOUBLE & Mean magnitude in BP-band (Vega)\\
180 & GAIA\_BP\_FLUX\_SNR & DOUBLE & Mean flux over standard deviation in BP-band (Vega) \\
181 & GAIA\_RP\_MAG & DOUBLE & Mean magnitude in RP-band (Vega)\\
182 & GAIA\_RP\_FLUX\_SNR & DOUBLE & Mean flux over standard deviation in RP-band (Vega) \\
183 & SDSS2GAIA\_SEP & DOUBLE & SDSS-Gaia separation in arcsec \\
\enddata
\end{deluxetable*}

\(\) %Text after forces full table to be displayed.

\end{document}